\documentclass[letterpaper,9pt]{sig-alternate}
\usepackage{epsfig,url}
\usepackage{times}
\usepackage{amssymb}
\usepackage{amsmath}
\usepackage{amsfonts}
\usepackage{graphicx}
\usepackage{color}
\usepackage{listing}
\usepackage{multirow}
\usepackage{float}
\usepackage{amssymb}
\usepackage{xspace}
\usepackage{subfigure}
\usepackage{color}
\usepackage{algorithm}
\usepackage{algpseudocode} 
\usepackage{bm}
\usepackage{ifthen}
\newboolean{report}

\let\oldbibliography\thebibliography
\renewcommand{\thebibliography}[1]{%
  \oldbibliography{#1}%
  \setlength{\itemsep}{-1pt}%
}

\usepackage[pdfpagelabels,
           plainpages=false,
           bookmarks=true,
           hypertexnames=false,
           pdffitwindow=false,
           colorlinks=true,
           linkcolor=blue,
           citecolor=blue,
           filecolor=black,
           urlcolor=black,
           pdfauthor={ Giuseppe Scavo, Zied Ben Houidi, Renata Teixeira, Stefano Traverso, Marco Mellia},
           pdftitle={WeBrowse: a Passive Content Curation System Based on HTTP Logs},
           pdfsubject={2015},
           pdfkeywords={media curation, HTTP traffic monitoring},
           pdfproducer={Latex with hyperref},
           pdfcreator={pdflatex}
          ]{hyperref}

\newcommand{\specialcell}[2][c]{%
  \begin{tabular}[#1]{@{}c@{}}#2\end{tabular}}

\long\def\comment#1{}

\newcommand{\HotNet}{W\textsc{e}\-B\textsc{rowse}\xspace}
\newcommand{\HTTPtrace}{\emph{ISP-week}}
\newcommand{\HTTPtracea}{\emph{ISP-PoP1-1day}\xspace}
\newcommand{\HTTPtraceb}{\emph{ISP-PoP2-1day}\xspace}
\newcommand{\HTTPtracec}{\emph{ISP-PoP3-1day}\xspace}
\newcommand{\UMB}{\emph{ISP-PoP3-1day}\xspace}
\newcommand{\POLITOa}{\emph{Campus-day-1}\xspace}

\newcommand{\GT}{\emph{HTTP-Alexa}}
\newcommand{\NT}{\emph{HTTP-GNews}}

\definecolor{red}{rgb}{1,0,0}

\title{Technical report \\ WeBrowse: Mining HTTP logs online for network-based content recommendation }

\numberofauthors{3}

\author{
    \alignauthor
    Giuseppe Scavo\\
    \affaddr{Inria, France}\\
    \hspace{-0.3cm}\affaddr{Nokia, Bell Labs}\\
    \alignauthor
    Zied Ben Houidi\\
    \hspace{-0.3cm}\affaddr{Nokia, Bell Labs}\\
    \alignauthor
    Stefano Traverso\\
    \affaddr{Politecnico di Torino, Italy}\\
    \and
    Renata Teixeira\\
    \affaddr{Inria, France}\\     
    \alignauthor
    Marco Mellia\\
    \affaddr{Politecnico di Torino, Italy}\\
}

\begin{document}

\setboolean{report}{true}

\maketitle
\begin{abstract}

A powerful means to help users discover new content in the overwhelming amount of information available today is sharing in online communities such as social networks or crowdsourced platforms.
This means comes short in the case of what we call \textit{communities of a place}: people who study, live or work at the same place. Such people often share common interests but either do not know each other or fail to actively engage in submitting and relaying information. To counter this effect, we propose \textit{passive crowdsourced content discovery}, an approach that leverages the passive observation of web-clicks as an indication of users' interest in a piece of content. 
We design, implement, and evaluate \HotNet, a passive crowdsourced system which requires no active user engagement to promote interesting content to users of a community of a place. Instead, it extracts the URLs users visit from traffic traversing a network link to identify popular and interesting pieces of information. We first prototype \HotNet\ and evaluate it using both ground-truths and real traces from a large European Internet Service Provider.  
Then, we deploy \HotNet in a campus of 15,000 users, and in a neighborhood.
Evaluation based on our deployments shows the feasibility of our approach. The majority of \HotNet's users welcome the quality of content it promotes. Finally, our analysis of popular topics across different communities confirms that users in the same community of a place share common interests, compared to users from different communities, thus confirming the promise of \HotNet's approach.

\end{abstract}

\section{Introduction}\label{Introduction}

The amount of information available on the web today, and the fast rate with which new information appears, overwhelm most users, including knowledge workers who account for 80\% of employees in north America~\cite{ManagementInformationSystems} and whose daily job is to digest and transform information.
Search engines have solved part of this information overload, but searching for something specific is only a fraction of users' interactions with the web. 
For instance, knowledge workers spend around 60\% of their web time either browsing with no specific goal in mind or gathering information from known websites~\cite{sellen2002knowledge}. 

A powerful means to discover new content today is sharing among members of online communities. We identify three main types of online communities: social networks, crowdsourced, and enterprise-specific. When users follow each other in social networks, they create communities, which are the basis for sharing content. Facebook, Pinterest, or Twitter illustrate how social networks can be effective to be well-informed~\cite{Kwak:2010:twitter}. Other users participate in online crowdsourced communities. For example, systems such as Reddit or HackerNews rely on users to submit and vote on content to promote. 
Finally, corporations encourage  employees to engage in Enterprise2.0 platforms~\cite{mcafee2006enterprise} such as wikis and corporate social networks to facilitate content discovery.

To be effective, content discovery in communities needs two ingredients: social connections and user engagement. First, members with common interests need to be connected so that information can flow. Second, they need to engage in submitting and sharing content. Unfortunately, these ingredients are often missing in the case of what we call \emph{communities of a place}: people who live, study, or work in the same place. 
For example, people in the same neighborhood should have interests in local news, events or shops but they rarely know each other personally. Co-workers in an organization also have common interests. People in business divisions might be interested in knowing the hottest topics or future technologies in their R\&D department, and vice versa. Despite the resources enterprises put on content discovery platforms, only few employees engage in sharing content~\cite{friedman2014enterprise}. Data we obtain from a large corporation's social network shows that, on average, only 0.3\% of workers contribute daily, and 10\% monthly. In practice, even Internet-wide systems such as Reddit and Digg suffer because only a relatively small community of users actively contributes~\cite{GilbertCSCW13,digg-patriots}.  

In this paper, we propose \textit{passive crowdsourced content discovery} to address the problem of under-engagement and the lack of social connections in communities of a place. Passive crowdsourcing in general is an approach that leverages the efforts of users without their active participation~\cite{passcrowd}. A popular example is passively observing users' locations and speeds to predict commute times.
Applying it to our problem, we leverage the passive observation of web-clicks (i.e., the URLs users intentionally visit) as an indication of users' interest in a piece of content. We assume that a click is a good measure of interest, as users often have an idea of the type of content they are about to access (e.g., because they saw a preview or because a friend recommended it). Intuitively, the more users click on a URL, the higher the interest in the content on the corresponding page. Our approach is then to leverage the collective clicks in a community to automatically discover relevant content to promote to users of the community. 

This idea sounds appealing as it rallies the power of crowdsourcing without the need for user engagement. However, the relatively small number of users composing a community of a place might be an issue, as it might limit the crowdsourcing effect. For example, Reddit has millions of users contributing with 200,000 submissions per day~\cite{reddit-stats}, while communities of a place typically consist of few thousand users. Hence, the main question for us is whether passive crowdsourcing is feasible for communities of a place. In particular, the communities should have enough users browsing the web to allow to passively identify content to promote, and users must share common interests (e.g., click on the same URLs).


To answer this question, this paper designs and deploys \HotNet (Sec.~\ref{sec:overview}), the first passive crowdsourced content promotion system based on web-clicks. To implement passive crowdsourcing, one must be in a position to observe the aggregated web-clicks of the community. Luckily, in many communities of a place, users will connect to the Internet from the same network, such as, e.g., the campus/enterprise network or the network of a residential Internet Service Provider (ISP) in a neighborhood. Hence, we base our service on the passive observation of web traffic flowing through a network aggregation point. 
\HotNet (i) observes web packets flowing through a network link, (ii) passively extracts HTTP logs (i.e., streams recording the headers of HTTP requests), and (iii) detects and decides on-the-fly the set of URLs to show to users. 


\HotNet is deployed in a large university campus with 15,000 users and recently in a residential internet service provider in a point of presence connecting 20,000 subscribers.\footnote{The website for the campus deployment is public: \url{http://webrowse.polito.it/}.}  We use data from both deployments to study the feasibility of our approach. 
We summarize the main contributions below.

\noindent \textbf{1. We are the first to use the passive observation of web-clicks to boost content discovery in communities of a place.}\\
\noindent \textbf{2. We develop methods to extract URLs that are interesting to users from the noisy HTTP logs.}
HTTP logs contain lots of requests that browsers generate automatically (e.g., to fetch ads or pictures), which do not correspond to any URL users explicitly clicked. Moreover, not all web pages are interesting to users at large. For example, users may often visit their bank website, but this is not the type of content that one would suggest to friends. Sec.~\ref{sec-filtering} describes the heuristics to identify the set of URLs candidate for promotion. Web portals such as \url{youtube.com} or \url{nytimes.com} are popular URLs, but we want to identify the specific video people are watching, or the news article they are reading. Sec.~\ref{sec-filtering-portal} presents an online algorithm to distinguish content from portal URLs. Promoting content based on the observation of users raises privacy concerns. Sec.~\ref{sec:privacy_perceiving} designs a set of privacy-preserving promotion algorithms.\\
\noindent \textbf{3. We show the feasibility and interest of passive crowdsourcing for content discovery in communities of a place:} 
We show that passively observing approximately one thousand active users is sufficient to identify enough interesting content to promote. Then, according to a poll, almost 90\% of users find the content \HotNet promotes interesting. We finally analyze popular topics in different communities of a place and find that people in the same community of a place share common interests. In particular, people in the same neighborhood share common interests, but less common interests compared to people in the campus, and more compared to people in another city.

\vspace*{-0.2cm}
\section{WeBrowse Overview}\label{sec:overview}
\begin{figure}
	\centering
	\hspace*{-0.5cm}\includegraphics[trim={0 14cm 8cm 0},width=1.1\columnwidth]{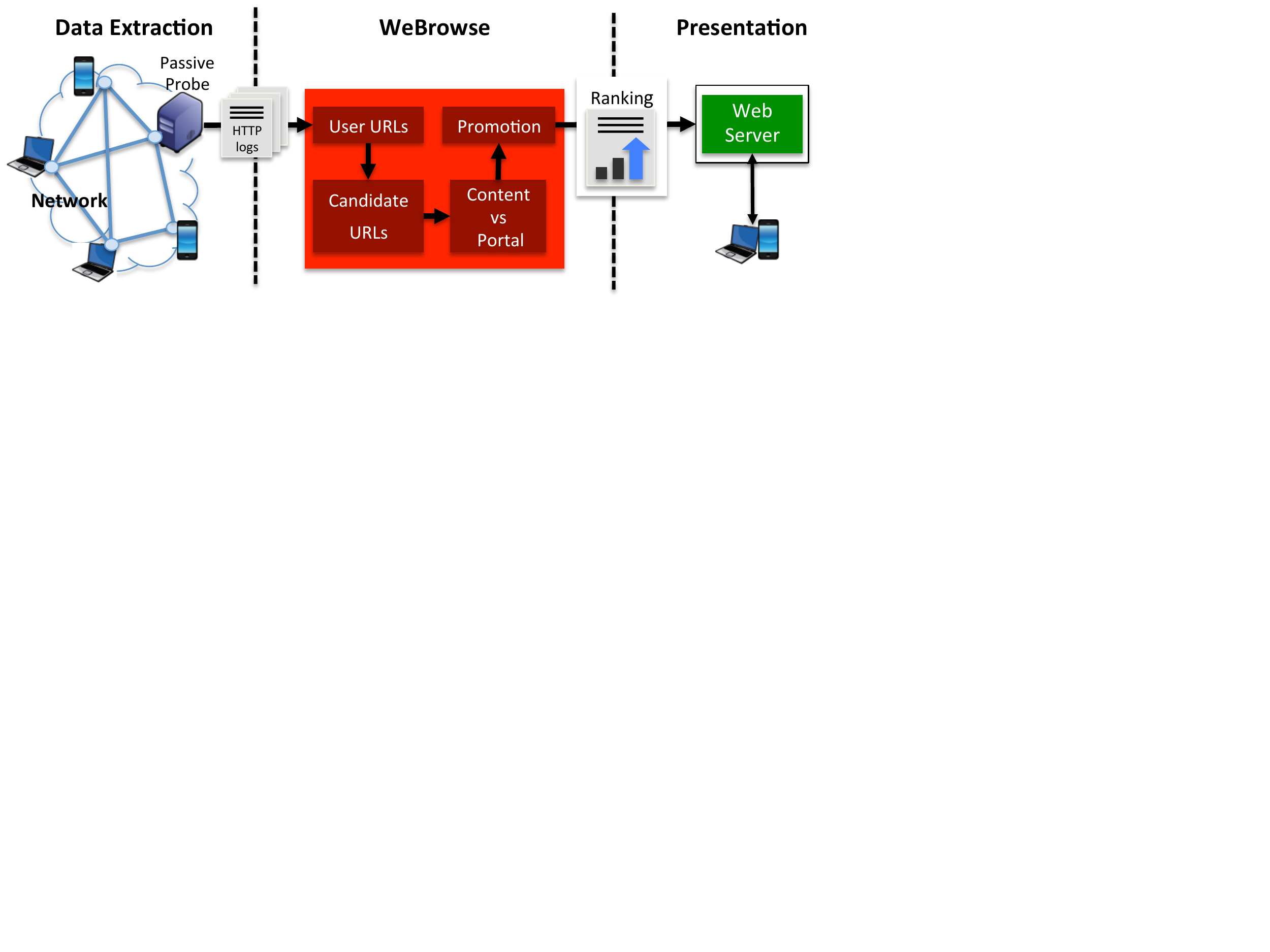}
	\vspace*{-0.4cm}
	\caption{An overview of the content promotion system.}
	\label{fig:cur:sys}
	\vspace*{-0.4cm}
\end{figure}

Fig.~\ref{fig:cur:sys} presents an overview of \HotNet. \HotNet\ takes as input HTTP requests from a raw data extraction module and outputs a list of sorted URLs to a presentation module, which is in our current deployments a website. 
Below, we describe the high-level architecture of our passive content curation service.

\noindent \textbf{Data extraction} is a a traffic monitor running within the network, it observes packets traversing a link and extracts HTTP requests.\footnote{Although HTTPS is gaining popularity~\cite{Naylor2014}, our analysis (not shown for conciseness) shows that only 7\% of websites \HotNet aims at promoting actually relies on encryption for data delivery. In a possible future when HTTPS will be dominant, we can still envision passive content curation based on alternative solutions, e.g., a browser plugin or corporate proxies that feed the passive promotion system.}  In this paper, we use the monitoring tool Tstat~\cite{finamore2011_tstat} for extracting HTTP requests, but this module can build on whatever packet capturing infrastructure ISPs already deploy (e.g.,~\cite{fraleigh:03}). Alternatively, we can extract HTTP requests from web proxies or even from plugins installed on user devices. We are interested in the following information from HTTP requests: \textit{timestamp}, \textit{URL} (obtained by chaining the information in \textit{host} and \textit{resource} fields), \textit{referer}, and \textit{user-agent} fields. We also extract an anonymized user identifier.\footnote{In Sec.~\ref{sec:privacy_perceiving}, we discuss why \HotNet needs this information and describe the techniques we adopt to preserve users' privacy.} All fields are extracted from the HTTP requests only, hence we ignore responses.
\\
\noindent \textbf{\HotNet} consists of four sub-blocks as depicted in Fig.~\ref{fig:cur:sys}. The \textit{user-URL filter} (Sec.~\ref{sec-filtering-browser}) identifies the HTTP requests corresponding to actual users' clicks, that we call \textit{user-URLs}. It eliminates the vast majority of HTTP requests that the browser automatically generates. The \textit{candidate-URL} (Sec.~\ref{sec-filtering-interest}) and the \textit{content versus portal} modules (Sec.~\ref{sec-filtering-portal}) together select the set of user-URLs that are worth sharing with other users. Finally, the \textit{promotion} module (Sec.~\ref{sec:privacy_perceiving}) takes as input this set of URLs (together with their timestamp) and decides which ones to output to the presentation module. 
\\ 
\noindent \textbf{Presentation} is similar to news aggregation and curation services. It takes as input the list of promoted URLs and presents them in a user-friendly web portal (similarly to Reddit).

\section{Datasets}\label{sec:dataset}

In this paper we use a set of traces, \textbf{ground-truth traces}, to validate \HotNet's modules and \textbf{HTTP logs}, to evaluate \HotNet. We summarize them in this section.
\\
\noindent \textbf{Ground-truth traces.}
We generate HTTP logs in a fully-controlled testbed similar to previous work~\cite{IMC-2011-referrer,xie-resurf-2013}. We manually visit the top-100 most popular websites according to Alexa ranking. When we are in the main page of each of these sites, we randomly visit up to 10 links they reference. We collect all the visited URLs as they appear in the browser bar. In parallel, we capture all the HTTP requests. We call the resulting HTTP log \GT. This trace contains a total of 905 user-URLs, corresponding to 39,025 HTTP requests. We also build a similar trace, \NT, by visiting almost 1000 news sites (user-URLs) from Google News. This trace contains 68,587 HTTP requests.
\\
\noindent \textbf{HTTP logs.} For this study, we employ several anonymized HTTP traces we collect at the backbone link of the campus network of Politecnico di Torino and at residential networks from a large ISP. We collect the residential traces at three routers located in three cities in Italy. We obtain them by running Tstat~\cite{finamore2011_tstat}, a passive probe which processes live the stream of packets in the network, tracks all HTTP requests and anonymizes identifiers (i.e., IP addresses) using Crypto-PAn~\cite{cryptopan}.
Table~\ref{tab:desc-traces} summarizes all our traces.

\begin{table}[t!]
\caption{Details of the traces considered in this study.}
\centering
\resizebox{1.0\columnwidth}{!}{\centering
\begin{tabular}{c|c|c|c|c}
\textbf{Trace}  & \textbf{Network}   & \textbf{Period}      & \textbf{Users} & \textbf{HTTP requests} \cr\hline
\HTTPtrace\     & ISP				 & 13-20 Jun 2013 	& 65,577 		& 190M  \cr
\HTTPtracea\    & ISP				 & 14 Apr 2015 		& 13,238 		& 16M  \cr 
\HTTPtraceb\  	& ISP			 	 & 14 Apr 2015	        & 3,019	 		& 1M \cr 
\HTTPtracec\    & ISP				 & 14 Apr 2015 		& 18,462		& 25M  \cr 
\POLITOa\  & Campus			         & 14 Apr 2015 		& 10,787 		& 21M  \cr

\end{tabular}
}
\label{tab:desc-traces}
\vspace*{-0.6cm}
\end{table}

\section{Identification of User and\\ Candidate URLs}
\label{sec-filtering}

Webpages have become considerably more complex in the last decade~\cite{ButToN2013}. When a user visits a page, the browser first issues an HTTP request to the main URL. In this paper, we call this webpage, which the user explicitly visits, \textit{user-URL}. Webpages typically embed diverse types of objects such as HTML, JavaScript, multimedia objects, CSS, or XML files. Ajax, JavaScript, and Flash objects may dynamically fetch other objects. Each of them is referenced with another URL and may be served from different (maybe third-party) servers. Thus, the browser issues individual HTTP requests for each object. Although fetching all these objects is necessary to render the page, the user does not directly request these objects. We call all these URLs \textit{browser-URLs}. Take the example of the user visiting `Web page~\#1' in Fig.~\ref{fig:web-page1}. The user visits URL-1, which is a user-URL. The browser then requests the URLs of the four embedded objects. This example also shows the type of each requested object (e.g., the main object is an HTML file).

Take again the example in Fig.~\ref{fig:web-page1}. After visiting `Web page \#1', the user clicks on one of the links in the page  (in this example, URL-1c) to visit `Web page \#2'. The figure also presents the referer field of each HTTP request. The \textit{referer} field contains the URL of the page that originated the request. In this example, the referer of the HTTP request to URL-1c is URL-1.

To build \HotNet, our first objective is to to automatically mine user-URLs from the set of all browser-URLs that we observe in a log. Once user-URLs identified, we are interested in detecting the URLs that qualify to be promoted, and that we call \textit{candidate-URLs}. This section extends our earlier work~\cite{Houidi_TMA14} that identified, offline, user-URLs and candidate-URLs. In particular, we augment our heuristics and modify our algorithms to run online.

\vspace*{-0.2cm}
\subsection{Detection of user-URLs}\label{sec-filtering-browser}

\begin{figure}[t!]
	\centering
   \includegraphics[width=0.8\columnwidth]{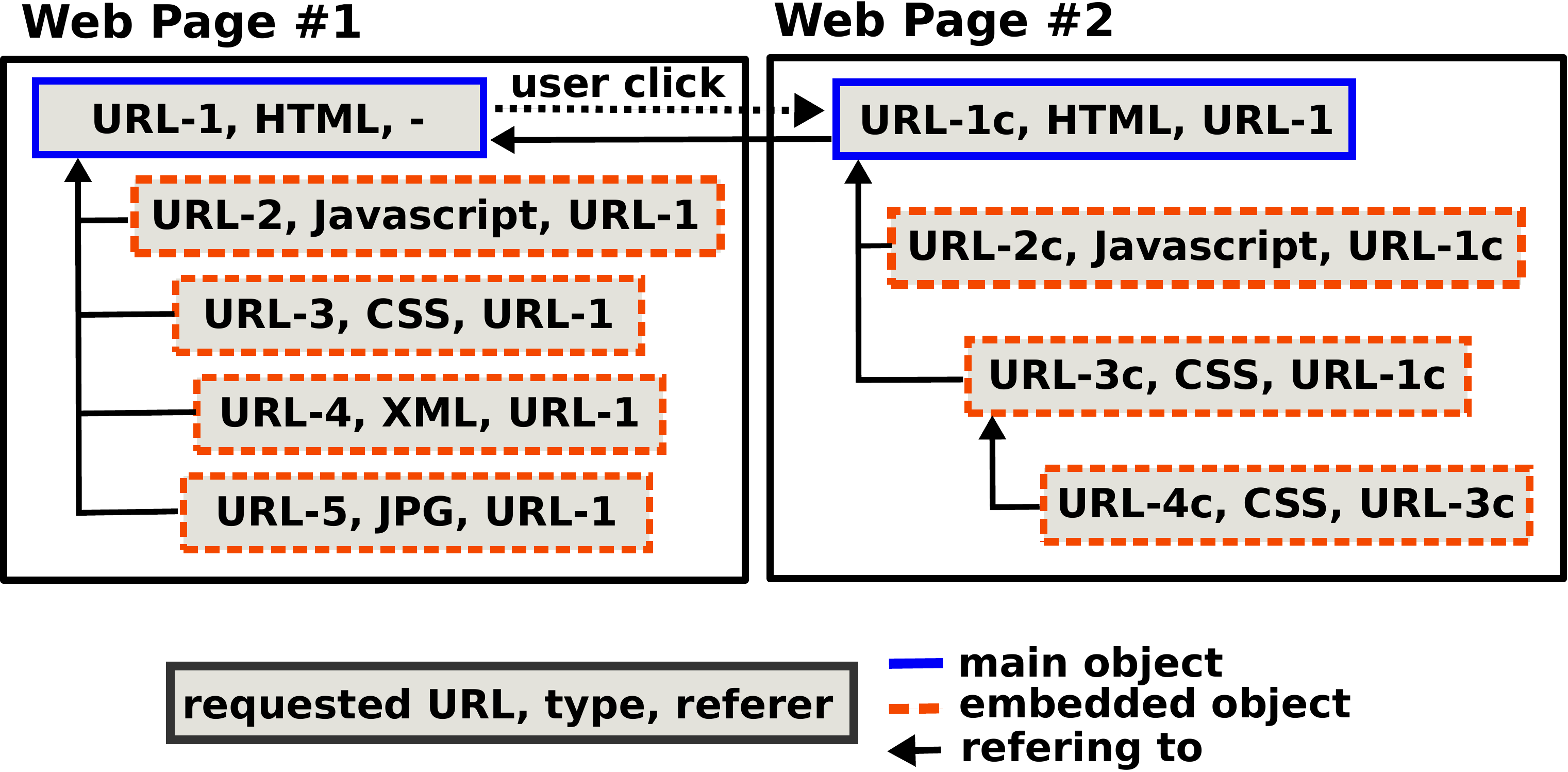}
    \vspace*{-0.4cm}
	\caption{HTTP requests for standard webpages.}
	\label{fig:web-page1}
	\vspace*{-0.5cm}
\end{figure}

We now illustrate our user-URL detection heuristics. We need to identify only the URLs issued by the users (e.g. Web page \#1', and four for `Web page \#2 in Fig.~\ref{fig:web-page1}). To this end, we first develop filters that build a candidate list of user-URLs from a set of HTTP logs. We have presented them in details in our prior work~\cite{Houidi_TMA14}, together with a comprehensive evaluation. We briefly recall them in the following.

\noindent \textbf{F-Ref.} It exploits the referer field in the requests to extract all URLs appearing as referer, e.g., URL-1, URL-1c and URL-3c in Fig.~\ref{fig:web-page1}. \\
\noindent \textbf{F-Children.} For each referer URL, say URL-1, it counts HTTP requests (children) with a URL-1 as a referer. URL-1 has five children in the example. F-Children removes from the candidate list URLs with less than $min\_c$ children. \\
\noindent \textbf{F-Type.} This filter checks the extension of objects in the URLs to discard those pointing to accessory files such as, e.g., \textit{.js}, \textit{.css}, \textit{.swf}. This filter would eliminate URL-3c. \\
\noindent \textbf{F-Ad.} It eliminates URLs pointing to known advertisement platforms using AdBlock's filter~\cite{ad-block}. \\
\noindent \textbf{F-Time.} It groups together all HTTP requests that happen within a time window $T$. This heuristic discards, within the time window, all the URLs that come after the first candidate user-URL. \\
\noindent \textbf{F-UA.} It checks the \textit{user-agent} field exposed in HTTP requests to discard those generated by non browser applications (e.g., DropBox, Google Play Store).

\subsection{Detection of candidate-URLs}\label{sec-filtering-interest}
Next, we select, among user-URLs, those worth sharing, what we call \textit{candidate-URLs}. Users often visit URLs related to webmail or e-banking, but these are not the kind of content they would appreciate as a recommendation. Our goal is then to identify the URLs that \HotNet\ should promote. 
To detect candidate-URLs, we leverage the presence of social buttons in a webpage, since it is an explicit indication that a user may want to share its content. Hence, our approach is to passively inspect the HTTP requests in the logs to detect children URLs pointing to well-known social network buttons (e.g., Facebook Share) widely adopted in the Web.\footnote{List available at \url{www.retitlc.polito.it/finamore/plugins.txt}}

Our earlier results~\cite{Houidi_TMA14} show that this method labels as candidate 70.72\% of the URLs contained in \NT, the ground-truth of candidate URLs we built visiting webpages promoted by Google News. We find that the remaining $~$30\% of URLs corresponds to websites embedding custom social sharing buttons, which deceive our heuristic (e.g., YouTube). Although we can reverse-engineer some of these ad-hoc methods to improve the accuracy, we leave it for future work given the complexity of this task.

In addition, for our upcoming deployments in corporate and research institution networks, we are currently experimenting a new heuristic to capture the different nature of content these communities are interested in. In particular, we simply tag as candidate-URL any user-URL containing a sufficient amount of text and a title. 

Finally, to understand how \HotNet\ filters candidate-URLs, we apply the heuristics from the previous section together with the heuristics to identify candidate URLs on three days of \HTTPtrace. As expected, the user-URLs represent a tiny fraction of all observed URLs. 
Out of 190 million requests, we identify 6.5 million user-URLs. Among these, only 422,500 are candidate-URLs.
%


\begin{algorithm}[t!]
\scriptsize
\begin{algorithmic}[1]
{
\Statex \hspace{-.5cm} \textbf{Input:} $\mathbf{HS}$, $T_{O}, min\_c, max\_p$ \# HTTP Request Stream, Observation Time, and parameters for F\_chidren and F\_param
\Statex \hspace{-.5cm} \textbf{Output:} $\mathbf{IS}$ \# Candidate-URL Stream 		
\Statex \hspace{-.5cm} \# Init Candidate Cache 					
\State $\mathbf{C} \gets \varnothing$
\Statex \hspace{-.5cm} \# Read current HTTP request
\While{h in $\mathbf{HS}$}
	\State $h \gets$ timestamp, URL, referer, user-agent, UserID
	\Statex \hspace{0.3cm} \# Check user-agent and URL is different from the referer
	\If{\Call{is\_browser}{h.user-agent} and h.URL != h.referer}	 
		\Statex \hspace{0.7cm} \# If current referer is not in Candidate Cache
		\If{h.referer $\not \in \mathbf{C}$ }
			\Statex \quad\quad\quad\quad \# If it passes type-based filter
			\If{\Call{f\_type}{h.referer}}
				\Statex \quad\quad\quad\quad\quad\quad \# Add referer to the Candidate Cache
				\State \Call{add}{$\mathbf{C}$, h.referer, timestamp}
				\State \Call{get\_candidate\_URL}{$\mathbf{C}$}
			\EndIf
		\Statex \quad\quad\quad \# If h.URL is a valid child
		\Else \If {\Call{f\_children}{h.URL, $min\_c$}}

			\Statex \quad\quad\quad\quad\quad\quad \# Increment the number of children and look for social 
			\State \Call{update\_info}{$\mathbf{C}$, h.referer, h.URL}
			\State \Call{get\_candidate\_URL}{$\mathbf{C}$}
			\EndIf
		\EndIf
	\EndIf
\EndWhile
\Statex

\Function{get\_candidate\_URL}{$\mathbf{C}$}
	\Statex \hspace{0.4cm} \# Iterate all referers in the Candidate Cache
	\For{ r in $\mathbf{C}$}
		\Statex \hspace{0.8cm} \# Check $T_{O}$ expiration and if it pass candidate filter
		\If{observation\_time(r) $> T_{O}$ and \Call{has\_social\_plugins}{r} and \Call{f\_children}{r, $min\_c$} and \Call{f\_params}{r, $max\_p$}}  
				\Statex \hspace{1.2cm} \# Send to the output
				\State write(r, $\mathbf{IS}$)	
				\Statex \hspace{1.2cm} \# Clean structures	
				\State remove(r, $\mathbf{C}$)
		\EndIf
	\EndFor
\EndFunction
\Statex

\Function{update\_info}{$\mathbf{C}$, referer, url}
	\Statex \quad\quad \# Increment the number of children
	\State \Call{increment}{$\mathbf{C}$.refer.children}
	\Statex \quad\quad \# Check the presence of social plugins
	\If{\Call{is\_social}{url}}
	\State \Call{set\_social\_plugin}{C.refer}
	\EndIf
\EndFunction
}

\end{algorithmic}
\caption{Online candidate-URL detector.}
\label{alg:interesting-online}
\end{algorithm} 

\subsection{Online algorithm}\label{sec:online:algo}
We design an algorithm that allows us to combine the heuristics from Sec.~\ref{sec-filtering-browser} and Sec.~\ref{sec-filtering-interest} to build several filtering configurations and process HTTP requests online. 

Alg.~\ref{alg:interesting-online} describes a simplified version of the algorithm that extracts candidate-URLs out of HTTP logs. It takes as input a stream of HTTP logs, $\mathbf{HS}$. It gets four fields for each HTTP request: \textit{$<$timestamp, URL, referer, user-agent$>$} and it returns a stream of candidate-URLs, $\mathbf{IS}$, in the format \textit{$<$timestamp, URL, referer$>$}.
This algorithm employs a hash table --  the \emph{Candidate Cache}, $\mathbf{C}$ -- which stores historical information for every observed referer for a limited \textit{Observation Time}, $T_O$.
As HTTP requests arrive, we keep only those with the user-agent in the browser white-list according to \textbf{F-UA} (line 4).
Then, we extract the referer, $r$, and we check its presence in $\mathbf{C}$. If $r$ is not in $\mathbf{C}$, we check the nature of its content with the \textbf{F-Type} filter (line 6). If $r$ passes the filter, we add it to the candidate cache $\mathbf{C}$ (line 7), which for a period of time equal to $T_{O}$ will store the timestamp of the first request having $r$ as referer, the number of children of $r$, and the social flag which is set to true if we observe a social plugin child for $r$. Conversely, if $r$ is in $\mathbf{C}$, we keep updated such information with \texttt{update\_info} (line 12).

Finally, we call the function \texttt{get\_candidate\_URL} (at lines 8 and 13).
For each referer in $\mathbf{C}$, we check if its observation time $T_O$ has expired. If so, it means that we can run the heuristics to label the URL as user-URL (for example, if the number of children is larger than threshold $min\_c$) and as candidate-URL if the flag signalling the presence of social plugins is true (line 20).
If the referer is candidate, we return it to the candidate-URL output (line 21).
Note that this algorithm can also output user-URLs.

\subsection{Accuracy}\label{sec:online:perf}
We evaluate the accuracy of our user-URLs detection algorithm, and we compare it with the best performing state of the art. For the evaluation we employ the \GT\ dataset and use two metrics:
(i) \textit{Recall}, i.e., the number of true positives (or URLs correctly labeled as user-URLs) divided by the number of positives (which is the total set of user-URLs). (ii) \textit{Precision}, i.e., the number of true positives divided by the number of true positives plus false positives (which is the number of URLs our heuristics say are user-URLs).

In our experiment, we compare different combinations of filters and different parameter tunings. 
Tab.~\ref{tab:perf:online} presents results for the most accurate filters\footnote{We exclude from this table the URLs to webpages delivered with HTTPS protocol.}. This table also compares with ReSurf~\cite{xie-resurf-2013}, the state of the art to detect user-URLs out of HTTP logs. ReSurf also leverages the referer field to separate HTTP transactions into multiple streams, but relies in addition on the HTTP \textit{content-type}, extracted from responses.


First, we observe that F-Ref + F-type + F-Children~($2$) + F-Param~($0$) filters when running online can achieve the same accuracy as ReSurf, if not slightly better (82.97\% of recall and 90.52\% of precision). More importantly, our algorithm is lightweight, and,
when comparing the processing time on the same trace, we find that our algorithm is around 25 times faster than ReSurf.
Second, our results show that there is a tradeoff between recall and precision. Indeed, F-Ref~+~F-type~+~F-Children~($2$)~+~F-Param~($0$) increases the precision (removing false positives) but comes at the cost of lower recall (it fails to detect some user-URLs).

Fortunately, we find that applying the candidate-URL filter on top of the user-URL filters has a positive side effect: it increases the user-URL detection precision to 100\%. For this reason, for our final implementation of \HotNet, we choose the filter with the best recall, F-Ref~+~F-type, and let the candidate-URL filter remove the remaining false positives.

Finally, we run additional tests to see the impact of increasing the observation time $T_O$, from $5$ seconds upto $30$ seconds. We find the best trade-off for $T_{O}=15$ seconds, even if this choice has very limited impact. We omit results for brevity. 

\begin{table}
 
 \caption{Performance and processing time achieved by ReSurf and our best performing combinations of heuristics for the detection of user-URLs on \GT\  (44686 requests).}\label{tab:perf:online} 
 \scalebox{0.75}{
 \centering
 \begin{tabular}{c|c|c|c}
   \textbf{Method} & \textbf{Recall} & \textbf{Precision} & \specialcell{\textbf{Processing time}} \\
   \hline
 ReSurf & 82.37\% & 90.33\% & 32484ms \\ 


\multirow{2}{*}{\parbox{5.6cm}{\centering F-Ref~+~F-type~+~F-Children($2$)~+~F-Param($0$)}} & \multirow{2}{*}{82.97\%}  & \multirow{2}{*}{90.52\%}  & \multirow{2}{*}{1270ms} \\
& & & \\
\textbf{F-Ref + F-type}  & 96.51\% & 63.60\% & 1251ms \\ 
  \end{tabular}}

\vspace{-0.1cm}
 \end{table}

\section{Content-URLs versus Portal-URLs}\label{sec-filtering-portal}

This section describes how \HotNet\ distinguishes candidate-URLs pointing to web portals from those pointing to specific content. We use the term \textit{web portal} (or \textit{portal-URL}) to refer to the front page of content providers, which generally has links to different pieces of content (e.g., \url{nytimes.com/} and \url{wikipedia.org/}); whereas a \textit{content-URL} refers to the webpage of, e.g., a single news or a Wikipedia article.
We first design an offline classifier. We then engineer an online version.

\subsection{Classification Features}
Given the heterogeneity of the URL characteristics, we opt for a supervised machine learning approach to build a classifier. We choose the Naive Bayes classifier, since it is simple and fast, thus, suitable for online implementation. As we will see, it achieves good performance, not calling for more advanced classifiers. 

We use five features to capture both URL characteristics and the arrival process of visits users generate.
\\
\noindent \textbf{URL length.} It is the number of characters in the URL. Intuitively, portal-URLs tend to be shorter than content-URLs.
\\
\noindent \textbf{Hostname.} This is a binary feature. It is set to one if the resource in the URL has no path (i.e., it is equal to ``/''); and to zero, otherwise. Usually, requests to portal-URLs have no path in the resource field.
\\
\noindent \textbf{Frequency as hostname.} This feature counts the number of times a URL appears as root of other candidate-URLs. The higher the frequency, the higher the chances that the URL is a portal.
\\
\noindent \textbf{Request Arrival Process (RAP) cross-correlation.} The URL request arrival process is modeled as a vector in which each element is the number of visits in five-minute bins. We notice that users often visit portal-URLs in a diurnal periodic pattern.
Intuitively, the more the request arrival process signal of a given URL is ``similar'' to that of a well-known portal, the higher the chances that the URL corresponds to a portal. To capture such a similarity we employ the cross-correlation, a well-known operation in signal processing that measures how similar two signals are, as a function of a sliding time lag applied to one of them. We compute the maximum cross-correlation between (1) the request arrival process of a tested URL and that of (2) well-known portals (e.g., \url{www.google.com} or \url{www.facebook.com}). The higher the value of the maximum of the cross-correlation, the larger the chance of a URL being a portal.
\\
\noindent \textbf{Periodicity.} This is a binary feature that builds on the observation that users visit portals with some periodicity. We use the Fast Fourier Transform (FFT) on the discretized aggregate visit arrival process for a given URL. If the visit arrival process shows one-day periodicity (principal frequency of 1/24h), then we set periodicity to one; zero, otherwise.

Some of these features require accumulating information in the system and pose practical constraints for an online implementation, which we address in Sec.~\ref{sec:online:algo:portal}. 

At last, note that some features work only for portals that are popular enough to observe enough clicks to detect the periodic diurnal cycle. Given that our promotion mechanisms (described in Sec.~\ref{sec:privacy_perceiving}) only promote URLs that are somewhat popular, it is unlikely that \HotNet\ will present to users URLs of unpopular portals.

\begin{table}[t!]
\begin{center}
\caption{Accuracy of the web portal/content classifiers.}\label{tab:perf:content:portal} 
\hspace*{-0.5cm}
\scalebox{0.65}{
\begin{tabular}{c|c|c|c|c|c|c|c}
  \hline
  \hline
  \multicolumn{2} {c} {\multirow{2}*{\textbf{Features}}} & \multicolumn{2} {|c|}{\textbf{Portal}} & \multicolumn{2} {|c|}{\textbf{Content}}  & \multicolumn{2} {c}{\multirow{2}*{\textbf{Accuracy}}}\\
  \cline{3-6}
  \multicolumn{2} {c|} {} & \textbf{Precision} & \textbf{Recall} & \textbf{Precision} & \textbf{Recall} & \multicolumn{2} {c}{}\\  
  \hline
  \hline
  \multicolumn{2} {c|} {Hostname}& 100\% & 23\% & 55\%  & 100\% & \multicolumn{2} {|c} { 60\%}       \\
  \hline
  \multicolumn{2} {c|} {Hostname +}& \multirow{2}*{100\%} & \multirow{2}*{75\%} & \multirow{2}*{80\%} & \multirow{2}*{100\%} & \multicolumn{2} {|c} {\multirow{2}*{87\%}}  \\
  \multicolumn{2} {c|} {RAP Cross-correlation}&&&&& \multicolumn{2} {|c} {}       \\
  \hline
  \multicolumn{2} {c|} {Hostname +}& \multirow{2}*{100\%} & \multirow{2}*{82\%} & \multirow{2}*{84\%} & \multirow{2}*{100\%} & \multicolumn{2} {|c} {\multirow{2}*{91\%}}  \\
  \multicolumn{2} {c|} {Periodicity}&&&&& \multicolumn{2} {|c} {}       \\
  \hline
  \multicolumn{2} {c|} {Hostname +}& \multirow{3}*{100\%} & \multirow{3}*{87\%} & \multirow{3}*{89\%} & \multirow{3}*{100\%} & \multicolumn{2} {|c} {\multirow{3}*{93\%}}  \\
  \multicolumn{2} {c|} {Periodicity +}&&&&& \multicolumn{2} {|c} {}       \\
  \multicolumn{2} {c|} {RAP Cross-correlation}&&&&& \multicolumn{2} {|c} {}       \\
  \hline
  \hline 
  \multicolumn{2} {c|} {URL length +}& \multirow{2}*{100\%} & \multirow{2}*{76\%} & \multirow{2}*{80\%} & \multirow{2}*{100\%} & \multicolumn{2} {|c} {\multirow{2}*{87\%}}  \\
  \multicolumn{2} {c|} {RAP Cross-correlation}&&&&& \multicolumn{2} {|c} {}       \\
  \hline
  \multicolumn{2} {c|} {URL length +}& \multirow{2}*{100\%} & \multirow{2}*{88\%} & \multirow{2}*{88\%} & \multirow{2}*{100\%} & \multicolumn{2} {|c} {\multirow{2}*{93\%}}  \\
  \multicolumn{2} {c|} {Hostname}&&&&& \multicolumn{2} {|c} {}       \\
  \hline
  \multicolumn{2} {c|} {\textbf{URL length +}}& \multirow{2}*{\textbf{100\%}} & \multirow{2}*{\textbf{93\%}} & \multirow{2}*{\textbf{94\%}} & \multirow{2}*{\textbf{100\%}} & \multicolumn{2} {|c} {\multirow{2}*{\textbf{96\%}}}  \\
  \multicolumn{2} {c|} {\textbf{Periodicity}}&&&&& \multicolumn{2} {|c} {}       \\
  \hline
  \hline
  \multicolumn{2} {c|} {All features }& 100\% & 93\% & 94\% & 100\% & \multicolumn{2}{c} { 96\%}  \\
  
  \hline
  \hline
  
 \end{tabular}
}
\end{center}
\vspace*{-0.6cm}
\end{table}

\subsection{Feature selection}\label{accuracy-content}
We build a dataset to train and test the accuracy of classifiers based on different features from \HTTPtrace. We manually visit 
each candidate-URL extracted from \HTTPtrace\ in order of appearance to label each as content-URL or portal-URL.  We perform this task until we get 100 URLs of each type. Note that to overcome class imbalance, we follow the approach of oversampling the minority class, i.e., portal-URLs~\cite{class-imbalance}. Then, we randomly divide the resulting 200 URLs into two sets: two thirds of the URLs for training and one third for testing. We use a ten-fold cross-validation, averaging results from 10 independent runs.


Tab.~\ref{tab:perf:content:portal} presents the accuracy of classifiers based on different combinations of features.
The table shows the recall and the precision in identifying portal-URLs and content-URLs, as well as the overall accuracy of the classifier. Identifying portal-URLs is an easy task: all combinations achieve a 100\% precision. However, lots of content-URLs are misclassified as portal-URLs. Interestingly, the hostname feature alone miserably fails to discriminate 
between portal-URLs and content-URLs. This is due to the fact that a lot of portal-URLs do not match their hostname (e.g. \url{www.nytimes.com/politics/}). Adding RAP and/or periodicity we enhance the performance, achieving
100\% precision and 93\% recall for identifying web portals, and more important, in our case: 94\% precision and 100\% recall for identifying content-URLs. This means that when using this combination we correctly label all the portals as portal-URLs, and we have a small probability (6\%) that the classifier wrongly labels portal-URLs as content-URLs.
At last, we obtain the same performance when considering all five features, suggesting that frequency, hostname and RAP do not bring more information. 


When engineering the online system we consider both the combinations URL length and hostname, and URL length and periodicity. The first combination allows an immediate decision, while the latter requires the system to collect information for some days.

\vspace*{-0.2cm}
\subsection{Algorithm for online classification}\label{sec:online:algo:portal}
The online algorithm we use to distinguish content-URLs from portal-URLs in real time takes as input a stream of tuples $<$candidate-URLs, timestamp$>$ and labels them  as content-URLs/portal-URLs. The algorithm sends URLs labeled as content-URLs to the promotion module, which will decide which ones to present to users.

The best performing feature combination, URL length and periodicity, requires us to collect in a database the timestamps of requests to candidate-URLs for some days. Therefore, we engineer the workflow of the algorithm to rely on this combination as soon as enough observations are available, and to use an on-the-fly classifier as a temporary solution. For the latter, we choose the best performing among the combinations that can execute on-the-fly, i.e., URL length and hostname. 

\begin{figure}[t!]
	\centering
    \includegraphics[trim=0cm 1cm 3cm 0cm, width=0.8\columnwidth]{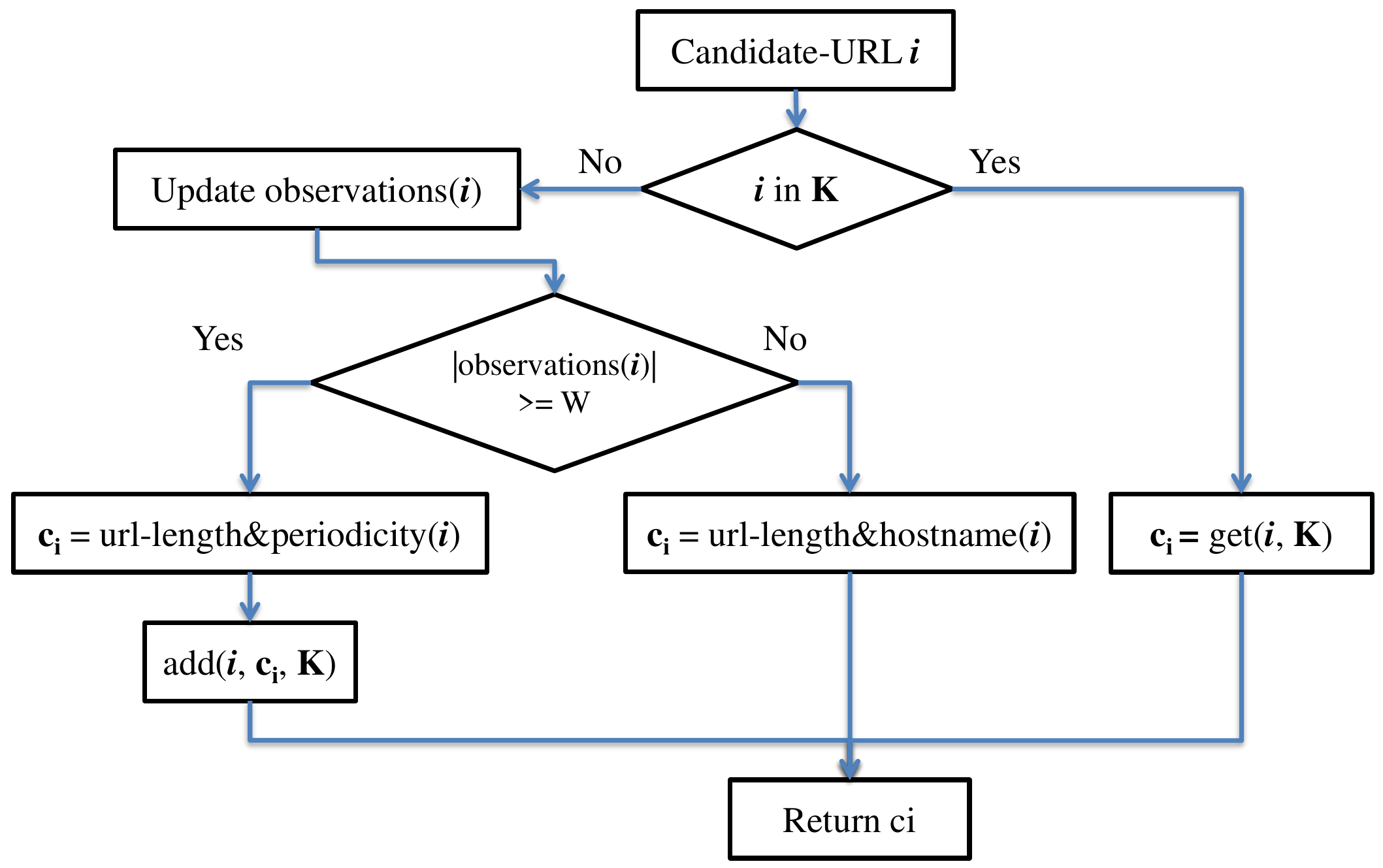}
	\caption{Workflow of the algorithm for online content vs portal URL classification.}
	\label{fig:online:classifier}
	\vspace*{-0.4cm}
\end{figure}

Fig.~\ref{fig:online:classifier} depicts the workflow of the algorithm. We employ the \textit{Knowledge Database}, $\mathbf{K}$, which the algorithm populates with the portal-URLs obtained by running the more precise classifier based on (url-length, periodicity). As soon as the algorithm receives a new candidate-URL $i$, it checks its presence in $\mathbf{K}$. If present, it immediately returns the classification result tagging the URL as a portal-URL. Otherwise, the algorithm counts the number of observations it has collected for $i$. If the number of observations is large enough ($\geq W$), the algorithm classifies $i$ using the classifier based on (url-length, periodicity), and stores the outcome $c_i$ in $\mathbf{K}$ for future decisions. Otherwise, the algorithm performs the classification on-the-fly using the classifier based on (url-length,hostname).

The accuracy of this method depends on the observation period of each URL. If the candidate-URL is new to the system we achieve the precision of the classifier based on URL length and hostname (88\%). It increases up to 94\% when enough information to run the classifier based on URL length and periodicity is available. 

To get an idea about the number of content-URLs in practice, we apply our algorithm on the same three days of \HTTPtrace{}  used in Sec.~\ref{sec-filtering-interest}. Out of the initial 90M distinct URLs, only 0.28\% (260,327) correspond to content-URLs.

\vspace*{-0.2cm}
\section{Privacy-preserving Promotion}
\label{sec:privacy_perceiving}
Once \HotNet\ identifies content-URLs, it must decide which ones to promote. Inspired by Reddit, we design content promotion modules based on both popularity and freshness. 
These two metrics are easily measurable from a network perspective.
In addition, we separate content-URLs according to their type: news, blog, video, and other content. We detect news using a predefined list of traditional news websites. For our deployment in Italy, we construct this list by crawling, every 20 minutes for a period of one week, the hostnames associated to news appearing on the frontpage of the Italian edition of Google News. This crawling allows us to obtain more than 500 distinct news websites that are indexed by Google News. For videos, we create a list with the most locally popular video sites (e.g., YouTube, Dailymotion, and video sections of popular newspapers). For blogs, we crawl the Italian blog index 
to get the list of the top-30k most popular ones. Finally, we label the content-URLs which do not fall in these categories as other content.  

\vspace{1.5mm} \noindent \textbf{Privacy threats} One of the most important requirements for passive crowsdourcing is to protect users' privacy. For \HotNet, we identify, in collaboration with the privacy and security boards of our institutions, three main privacy threats that are related to content promotion (we conduct a similar analysis for the data collection part): i) Personal information can leak into promoted URLs. For example a URL of a website showing users' login credentials. We design promotion mechanisms that prevent this threat. ii) Private pages with open access could be promoted. For example, \HotNet could promote the URL of a hosted private image. Luckily, such URLs do not fall under our definitions of candidate-URL. 
iii) Inferring personal information by combining \HotNet's data with information from other sources (e.g. list of persons present on site, or particular preferences of certain users). 
The risk analysis conducted by the boards labeled this latter threat as very improbable.


\vspace{1.5mm} \noindent \textbf{Promotion}
Having in mind above issues, we build three promotion modules in the \HotNet's website. \\
\noindent $\bullet$  \textit{Live Stream.} This module promotes web pages that are freshly clicked and belonging to news, videos, and blogs categories. \\
\noindent $\bullet$  \textit{Top.} This module promotes web pages by building a rank based on their number of clicks. In the current version, we show the ranking for different time spans: one day, one week, and one month. \\
\noindent $\bullet$ \textit{Hot.} This module builds on Reddit's Hot ranking algorithm~\cite{reddit-hot}, which promotes URLs that are both popular and recent. This algorithm assigns each content a score based on users' up and down votes. The links are then presented by decreasing score. For our case, we replace votes with the number of clicks, and modify the ranking score to obtain the following:
\vspace{-0.2cm}
\[
	Score(u)=log(N_{views}(u))+\frac{T_{first}(u)-T_{0}}{T_{P}}
\]
where, for each content-URL $u$, $N_{views}(u)$ reports the number of clicks, $T_{first}(u)$ is the timestamp corresponding to the first time $u$ has been observed, and $T_{0}$ is the timestamp corresponding to an absolute reference (i.e., the start of \HotNet deployment). Finally, $T_P$ is a normalization factor defining the ``freshness period''. We set it to 12 hours. 
Intuitively, this score finds a balance between the content popularity and its recentness (i.e. its age with respect to the absolute reference). URLs new to the system get larger $T_{first}$, and consequently, a larger second term of the formula, increasing their probability to reach the top rank positions because of their freshness, even if yet relatively popular. However, when a URL stops attracting attention (i.e., clicks), the first term of the formula stops increasing, and the URL is likely to get surpassed by fresher URLs.
Finally, the hot module keeps and permanently updates the scores and rankings of content-URLs. Content URLs are then presented to the users by decreasing score.

To protect users' privacy from the threats described above, we adopt the following techniques. For the case of Live Stream module, we protect users' privacy by \emph{(i)} promoting only content coming from public, trusted and known hostnames, and, as an additional security measure, \emph{(ii)} stripping all the parameters contained in URLs.
For the Top and Hot modules, we opt for $k$-anonymity, i.e., a content-URL is allowed to be promoted iff \HotNet has observed at least $k$ different userIDs (built by combining the IP address of the clients and the User-Agent field in the header of HTTP requests) accessing it. We choose $k$=5 and do not show URLs older than 24 hours. With this configuration, the same user has to visit the same URL, assuming it contains sensitive information, from five different devices in one day for her personal information to be exposed. 
Finally, we allow the users to opt-out from the system by enabling the DoNotTrack flag in their browsers, and ignoring all HTTP requests containing the flag. 
These policies have been discussed and agreed with the Security and Privacy boards for our deployments.

\vspace{1.5mm} \noindent \textbf{Removing \HotNet bias.} Finally, since our modules query promoted URLs to create previews, we have to remove the effect of \HotNet on itself. Similarly, content promoted on the portal has a higher probability of being visited by the users. Both of these artifacts inflate promoted URLs' popularity. To counter this effect, we instrument the system to ignore requests having \HotNet\ website as a referer.
\vspace*{-0.2cm}
\section{System Evaluation}\label{sec:online:performance}

In this section we show how our filtering works on real ISP traces and we briefly discuss the resource demand of our online algorithms when working in real-time.

\subsection{WeBrowse filtering}\label{sec:applying}

To understand how \HotNet\ filters content-URLs in practice we apply it on three days of \HTTPtrace.
The user-URLs represent a tiny fraction of all the observed URLs in the trace. 
Out of the 90M distinct URLs, only 0.28\% of them correspond to content-URLs.

Our method detects that among the 190 million a requests, only 3.4\% correspond to visits to actual user-URLs. Among these user-URL visits, only around 6.5\% of them are to interesting-URLs, corresponding to 0.22\% of the total number of visits. 
Among those the 3,191 detected portal-URLs cumulate 57,794 visits (18.1 per-portal visits on average over 3 days), and the 260,327 classified as content-URLs cumulated 373,147 visits (1.43 visits on average). 

These results highlight the challenge of filtering the humongous amount of traffic to pinpoint the URLs that are interesting. 

\subsection{Performance Evaluation}\label{sec:performance}

\begin{figure}[t!]
        \includegraphics[width=1.0\columnwidth]{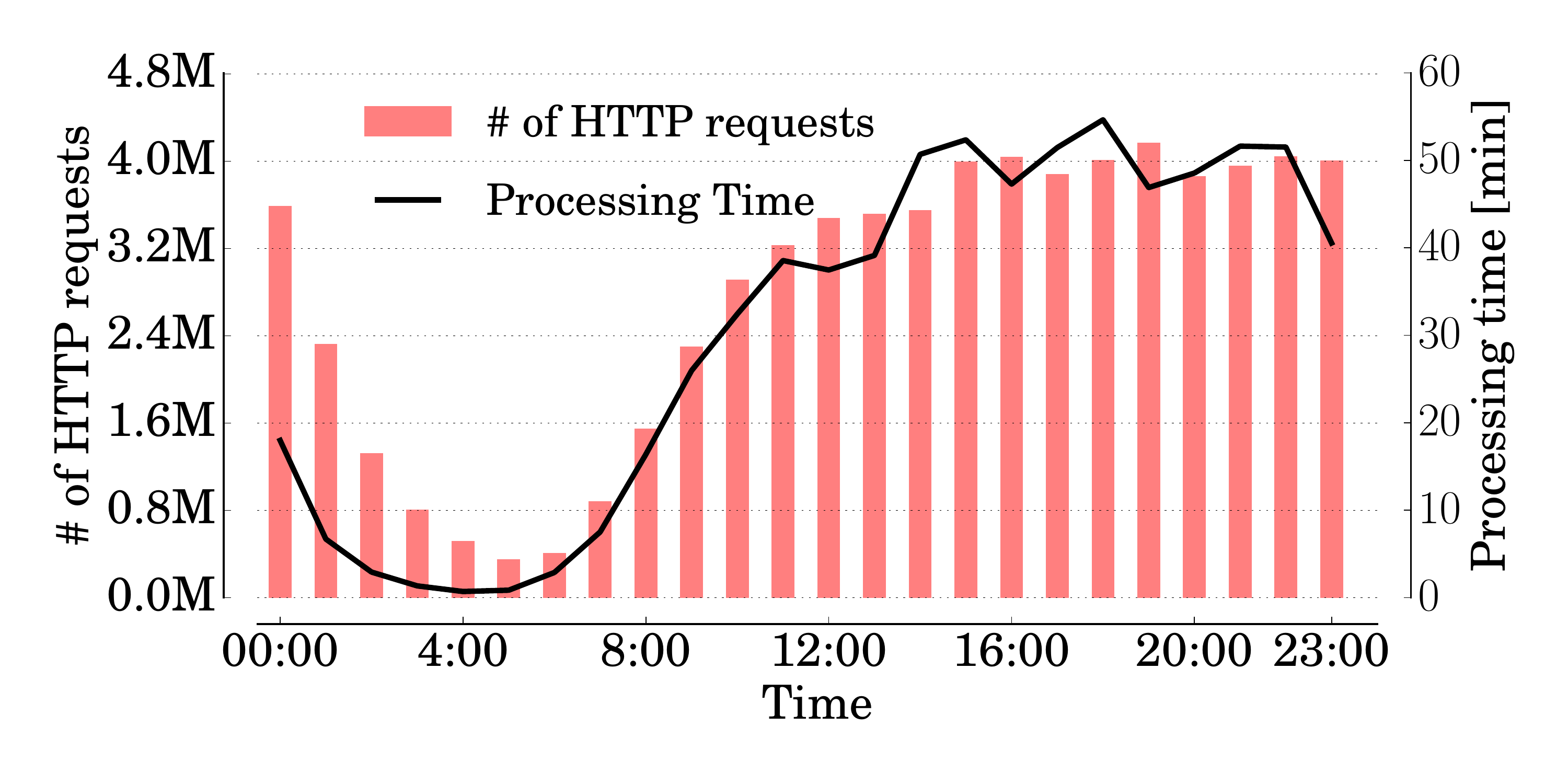}
        \vspace*{-0.8cm}
        \caption{HTTP requests rate (left y axis) and processing time (right y axis) over time for one day extracted from \HTTPtrace.}\label{fig:processing_time}
        \vspace*{-0.6cm}
\end{figure}

We evaluate the performance of our \HotNet\ implementation to process requests in \HTTPtrace, which aggregates the HTTP traffic from about 20,000 households.
We select the day in \HTTPtrace\ with the largest peak hourly rate of HTTP requests.
We split the one-day trace in 1-hour long sub-traces, and use them to feed \HotNet. For each hour, we measure the time that \HotNet\ spends to end the processing. 
For this experiment, we run \HotNet\ on a a server equipped with a 2.5GHz CPU and a 32GB RAM.
%
Fig.~\ref{fig:processing_time} reports the amount of HTTP requests (left-hand y axis) and the corresponding processing time (right-hand y axis), for all the 1-hour long bins in the day of \HTTPtrace. 
The figure shows that \HotNet's implementation is able to complete up to 4M HTTP requests in less than one hour. This demonstrates that \HotNet\ can sustain the processing of such large rates of HTTP requests on a rather standard server like the one we pick. Finally, we note that \HotNet's memory footprint is minimal, as less than 960MB of memory was used throughout the experiment.

We are convinced that the content promoted by \HotNet can only become more interesting with a larger population of users. When scaling to, e.g., a country-wide ISP network scenario, \HotNet\ will be asked to analyze the HTTP traffic of millions of users, which is challenging. 
Hence, to let the system scale, we could spatially distribute \HotNet deployments, or adapt \HotNet to work based on BigData paradigms. We leave the exploration of this aspects for future work.

\vspace*{-0.3cm}
\section{WeBrowse Deployments}\label{sec:online:deploy}
We have deployed an operative version of \HotNet in the campus network of Politecnico di Torino. Then, we publicly announced \HotNet to the campus community allowing us to report on users' feedback on the content \HotNet promotes. We have also installed \HotNet at an ISP PoP providing Internet access to a residential neighborhood of an Italian city. This deployment is still on trial. 

Each deployment uses two servers: The probe, running Tstat, extracts HTTP requests and streams them to \HotNet's backend server which extracts content-URLs and promotes a subset of them on the website. The campus promotion site is public at \url{http://webrowse.polito.it/}. 
The website features the three promotion tabs discussed in Sec.~\ref{sec:privacy_perceiving} and an additional one promoting contents from the websites belonging to the university's domain. 
Each tab contains a content feed inspired by the ``wall'' implemented in popular social networks such as Facebook and Twitter. URLs in the feed have a  preview image, a title, and a short description when available. 

\vspace{1mm} \noindent \textbf{Scalability}
To estimate the amount of resources needed for our backend servers, we ran first \HotNet offline on \HTTPtrace, which aggregates the HTTP traffic of about 20,000 households and approximately 65,000 users.
We measure the time that \HotNet\ spends to process the trace using a server with off-the-shelf hardware configuration (2.5GHz CPU and a 32GB RAM). 
\ifthenelse{\boolean{report}}{}{Our results (see~\cite{techrep} for details) show that \HotNet's implementation is lightweight enough to process up to 4M HTTP requests in less than one hour using a marginal memory footprint, and demonstrate that \HotNet\ can sustain the processing of such large rates of HTTP requests on a rather standard hardware configuration.}

Although the campus network aggregates approximately 15,000 users (students, professors, and staff), it generates less traffic compared to the residential ISP scenario.
For this deployment, we use a backend server equipped with a quad-core 2.66GHz CPU and a 4GB RAM, which we observe to be enough to sustain HTTP request rate for this scenario.
The \HotNet\ server processes in real-time HTTP logs containing on average 21M requests per day and extracts on average 156,000 user-URLs, corresponding to 35,000 candidate-URLs. Roughly, this represents around 18\% of daily submissions in Reddit, with 300 times less active users in our case.

\vspace{1mm}\noindent \textbf{Data protection.} 
As we did for the promotion methods, we must ensure our data collection respects users' privacy. To this end, we engineered the system so that the backend server receives the anonymized stream of HTTP logs from the probe, and keep userIDs only in RAM for 24 hours (for k-anonymity). Moreover, all data is stored in a secured server with restricted access. This procedure got the approval from the Security Office of Politecnico di Torino and the ISP (equivalent of an IRB approval in the United States).

\vspace{1mm}\noindent \textbf{Campus user study.} We advertised \HotNet broadly in two rounds (to different mailing lists of the university) reaching around 6,000 users in Politecnico di Torino. We asked users to provide us a feedback about \HotNet by filling the evaluation form available on the website. See Sec.~\ref{sec:online:feedbacks} for details.

\ifthenelse{\boolean{report}}{
\section{Content Evaluation}\label{sec:evaluation}
}{
\section{Evaluation}\label{sec:evaluation}}
\vspace{1mm} \noindent \textbf{Evaluation challenges}
Evaluating a system like \HotNet is very challenging: there is no ``perfect-match'' competitor to \HotNet in the campus community, and existing platforms are not really comparable. Reddit is not popular in Italy because only few users contribute to it. Our user poll (see Sec.~\ref{sec:online:feedbacks}) shows that users in the campus rely mainly on news media and Facebook to get informed. Comparing \HotNet to news media is not fair because news is only a small part of what \HotNet promotes \ifthenelse{\boolean{report}}{(see Sec.~\ref{sec:complement})}{(see Sec.~9.4 of our technical report~\cite{techrep} for details)}. What we do instead is to check whether the topics of the news that users consume in the campus match what the Italian version of Google News promotes (Sec.~\ref{sec:communities}). Analyzing Facebook feeds of users in the campus is not possible, because it is hard to get a representative number of volunteers. Finally, Twitter has no trending data based on the campus location.

Instead, we focus our evaluation on a number of objective and subjective metrics. The objective metrics will help us answering the challenges that come with \HotNet: What is the critical mass of users for \HotNet to work? (Sec~\ref{sec:liveness}) And is there really a community of a place effect? (Sec~\ref{sec:communities}). The subjective metrics consist in collecting users' feedback on \HotNet's content (Sec~\ref{sec:online:feedbacks}).

\subsection{Effect of the crowd size}\label{sec:liveness}

\HotNet depends on users accessing the web from the network to identify the content to promote. The more users on the network, the more dynamic \HotNet will be and the more clicks the promoted content will have. 
We measure, for our campus deployment, the \textit{liveness} of \HotNet with the number of new added links in the Hot tab, and study how many users are needed for \HotNet to be ``alive''.

Fig.~\ref{fig:liveness} presents three sub-plots reporting statistics collected every 30mins. The bottom plot presents the number of active users in the network, which we approximate by the number of active userIDs; the middle plot presents the total number of user-URLs; and the top plot reports the number of URLs which appear for the first time in the Hot tab (new Hot URLs). We observe a typical day/night pattern with the number of active users which grows during working hours and decreases during the night and weekends. The number of new Hot URLs presents a positive correlation with the number of active users. 
\ifthenelse{\boolean{report}}{In the plots we compare the data from the campus trace with data from ISP trace. We notice that in the campus trace there is not new Hot URLs during night and weekends. This is expected, as many users leave the area during these days. In the Campus, we rarely observe new Hot URLs when the number of active users is below 200. 
The Hot tab is clearly more dynamic when there are at least 800 active users in the network. As the users of the ISP trace are more active and produce more User-URLs, the Hot Tab requires less users to be dynamic. In the ISP trace we observe new Hot URLs starting from 100 users. The higher activity in the ISP trace is partly due to the large amount of internal-network URLs that people in the Campus consume and we do not take into account in this plot.}

{In particular, there are no new Hot URLs during nights and week-ends. In this deployment, we rarely observe new Hot URLs when the number of active users is below 200. 
The Hot tab is clearly more dynamic when there are at least 800 active users in the network. However, this scenario is challenging for \HotNet because of the relatively low browsing activity of users in the campus. In fact, results for the ISP deployment show that the number of active users needed to the Hot tab to be dynamic is even lower (see our technical report for details~\cite{techrep}).}



\ifthenelse{\boolean{report}}{
\begin{figure}[t!]
	 \centering
	\hspace*{-0.5cm}\includegraphics[width=1.1\columnwidth]{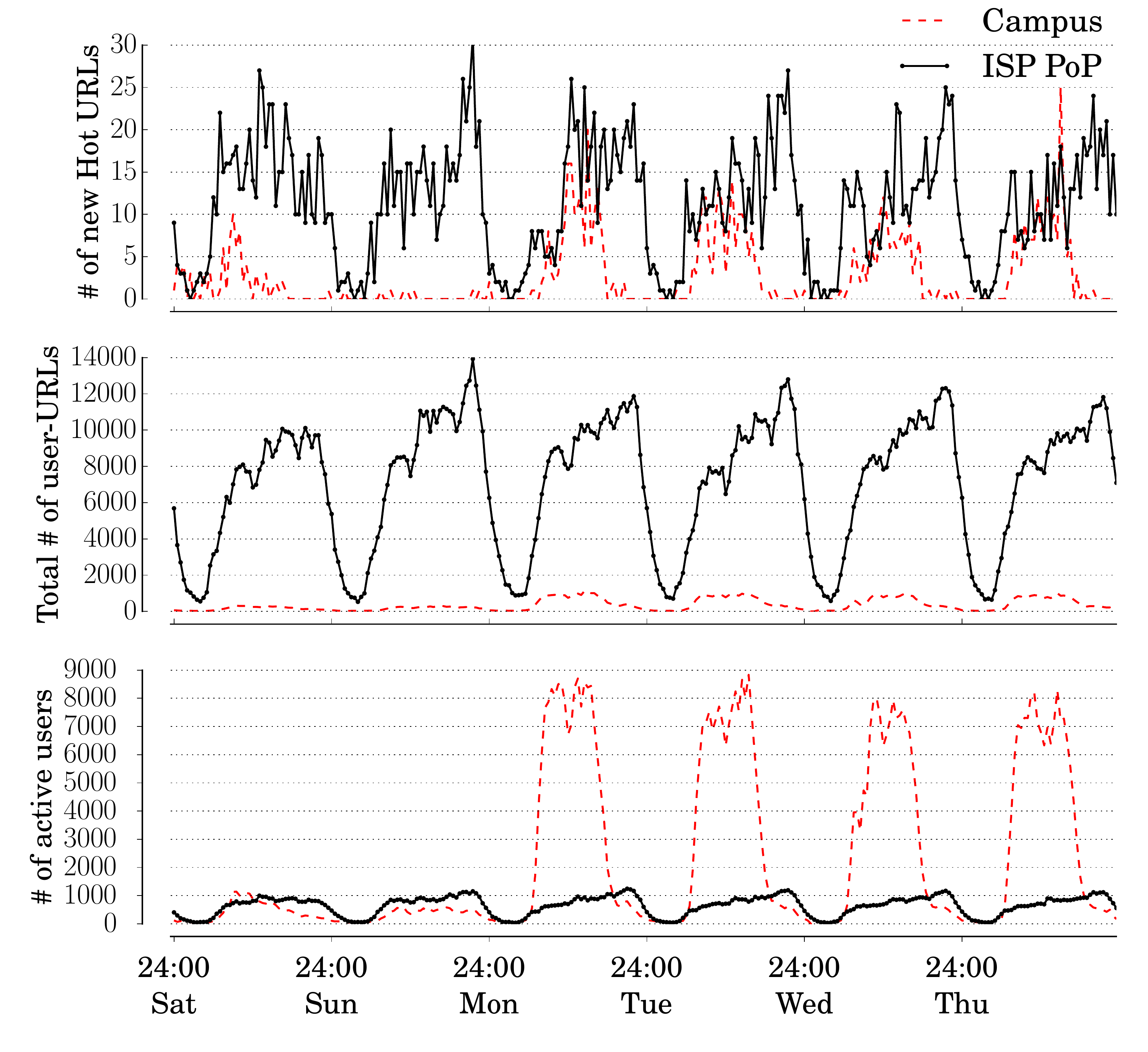}
	\vspace*{-0.8cm}
	\caption{The number of active users observed by \HotNet (bottom), and number of never seen before URLs in the Hot category (top). Statistics collected from \HotNet's campus deployment every 30min.}
	\label{fig:liveness}
	\vspace*{-0.6cm}
\end{figure}
}{
\begin{figure}[t!]
	 \centering
	\hspace*{-0.5cm}\includegraphics[width=1.1\columnwidth]{figures/liveness_poli_fw.pdf}
	\vspace*{-0.8cm}
	\caption{The number of active users observed by \HotNet (bottom), and number of never seen before URLs in the Hot category (top). Statistics collected from \HotNet's campus deployment every 30min.}
	\label{fig:liveness}
	\vspace*{-0.6cm}
\end{figure}
}

 
\subsection{User study}\label{sec:online:feedbacks}

We study the feedback we got form users visiting \HotNet's website and filling the evaluation form to understand whether they subjectively like \HotNet content. Second, we analyze users' interaction with the website by using Google Analytics. 
One might think that people in the campus are positively biased towards us. To counter this effect, we proceed to two rounds of announcements and user feedbacks collections. The first (R1 in short) targets a closer community of 300 students, professors and staff from the computer science department. The second round of advertising (R2) reached a wider population, i.e., professors, researchers, and students from different areas (engineering and architecture) and administrative employees.

One week after the email announcements, we observe visits coming from 1506 distinct users.
Interestingly, we observe that 20\% of users kept visiting the website after their first visit. The average visit is 2.5min long. Out of the users who visited the website, 115 ($~$8\%) filled our evaluation form.
Although the number of respondents is small, their feedback is extremely valuable for our future work on \HotNet.\footnote{The response rate we obtain is in line with that of surveys with no invite incentive~\cite{surveyrate}.}

We summarize the feedback of the 115 respondents in the following. We split the questions in our evaluation form into two main groups. The first group helps us evaluate whether users like \HotNet\ and the second focuses on our promotion methods. Not all respondents answered all questions.

\vspace{1mm} \noindent \textbf{Do users like \HotNet?}
We ask users questions about their experience with \HotNet\ and to compare it with the service they use to discover content on the web. Tab.~\ref{table:questionnaire} summarizes the responses. Overall, respondents were positive: the wide majority of respondents find \HotNet\ at least interesting or extremely interesting. 
Similarly, 71\% in R1 and 91\% in R2 of respondents find it useful or extremely useful. Interestingly, people from other departments were more positive than colleagues in the same department. 
More interestingly, responses during working hours (9am to 6pm) are more enthusiastic. This positively correlates with the dynamic behavior of \HotNet.

We also ask users in R2 to list the services they usually use to stay informed, and compare them to \HotNet. As shown in Tab.~\ref{table:questionnaire}, 25 of
respondents rely on news portals; Facebook comes second (12 respondents). Interestingly, 41 respondents say that \HotNet\ is simply different from these services. 
These answers are encouraging as we see \HotNet\ as a complement, and not a replacement, to existing systems. 

Finally, the 62\% (70\%) in R1 (R2) say they would like to have \HotNet as a service offered by their network. 44\% (30\%) in R1 (R2) would use \HotNet at least once a day.

\vspace{1.5mm} \noindent \textbf{How good are \HotNet's promotion algorithms?}
We ask users to rank the three different tabs in \HotNet\ (Top, Hot and Live Stream) using a Likert Scale from the most interesting to the least interesting. We calculate the average ranking score for each tab. 
The scores are fairly close, with the Top tab coming first, then Hot, and Live Stream as last. This result indicates that users have different tastes, and having different tabs with different promotion methods is important
to please a large user population.

\begin{table}
\begin{center}
\caption{User feedback from the evaluation form.}\label{table:questionnaire}
\vspace*{0.1cm}
\hspace*{-0.5cm}
\scalebox{0.65}{%
\begin{tabular}{c|l|l c c l l}
  \specialcell{\textbf{How interesting is} \\ \textbf{\HotNet's content?}} & R1 & R2 &  & \textbf{How useful is \HotNet?} & \multicolumn{1}{|l}{R1} & \multicolumn{1}{|l}{R2} \\
  \cline{1-3} \cline{5-7}
 extremely interesting  &  8  &  24  & & extremely useful   & \multicolumn{1}{|l}{4} & \multicolumn{1}{|l}{14}     \\
 very interesting       &  25 &  19  & & very useful        & \multicolumn{1}{|l}{20} & \multicolumn{1}{|l}{20}     \\
 interesting            &  17 &  10  & & useful             & \multicolumn{1}{|l}{18} & \multicolumn{1}{|l}{16}     \\
 poorly interesting     &  5  &  1   & & poorly useful      & \multicolumn{1}{|l}{10} & \multicolumn{1}{|l}{1}     \\
 not relevant           &  4  &  12   & & not relevant       & \multicolumn{1}{|l}{7} & \multicolumn{1}{|l}{4}     \\
    \multicolumn{4}{l}{}\\
\end{tabular}
}
\scalebox{0.65}{%
\begin{tabular}{c|l c c l}
 
   \specialcell{\textbf{Which service do you use} \\ \textbf{to keep informed?}} &  R2 & &  \specialcell{\textbf{How do you compare} \\  \textbf{this service to  \HotNet?}} & \multicolumn{1}{|l}{R2} \\
    \cline{1-2} \cline{4-5}
    Web Newspapers  &  25   & & More interesting   & \multicolumn{1}{|l}{5}   \\
    Facebook        &  12   & & Less interesting   & \multicolumn{1}{|l}{7}   \\
    Google News     &  4    & & Simply different   & \multicolumn{1}{|l}{41}   \\
    Twitter         &  3    & &                   &  \\
    Newsletters     &  2    & &                   &  \\
    Other Media     &  10   & &                   &  \\
\end{tabular}
}
\end{center}
\vspace{-0.6cm}
\end{table}



\begin{figure*}[t!]
\hspace*{-0.5cm}
\subfigure[Same Campus]{
    \includegraphics[width=0.21\textwidth]{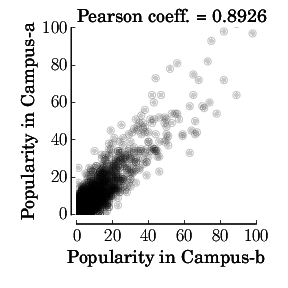}
    \label{fig:polito_polito}}
		\hspace*{-0.5cm}
\subfigure[Same city]{
    \includegraphics[width=0.21\textwidth]{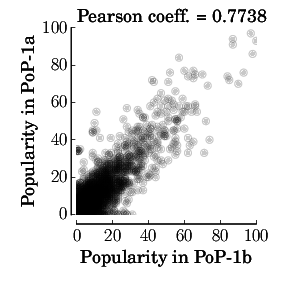}
    \label{fig:milano_milano}}
\hspace*{-0.5cm}
\subfigure[Campus vs. Residential]{
    \includegraphics[width=0.21\textwidth]{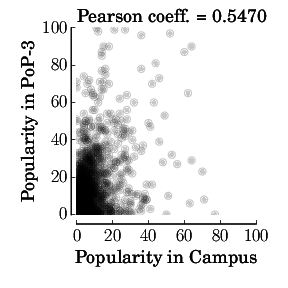}
    \label{fig:polito_torino}}
		\hspace*{-0.5cm}
\subfigure[Different cities]{
    \includegraphics[width=0.21\textwidth]{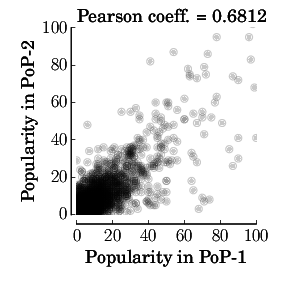}
    \label{fig:milano_torino}}
\hspace*{-0.5cm}
\subfigure[News]{
    \includegraphics[width=0.21\textwidth]{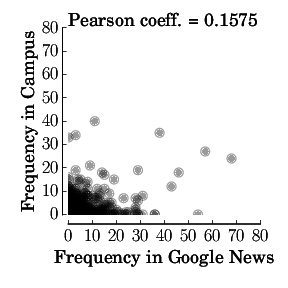}
    \label{fig:polito_gnews}
		}
\vspace*{-0.4cm}
\caption{Scatter plots where dots represent keywords extracted from different trace pairs and coordinates correspond to their popularities (Figs.~\ref{fig:polito_polito}-\ref{fig:milano_torino}) and their frequency in news webpages (Fig.~\ref{fig:polito_gnews}).}
\vspace*{-0.4cm}
\end{figure*}

\ifthenelse{\boolean{report}}
{\begin{figure}[t!]
	 \centering
	\includegraphics[scale=0.7]{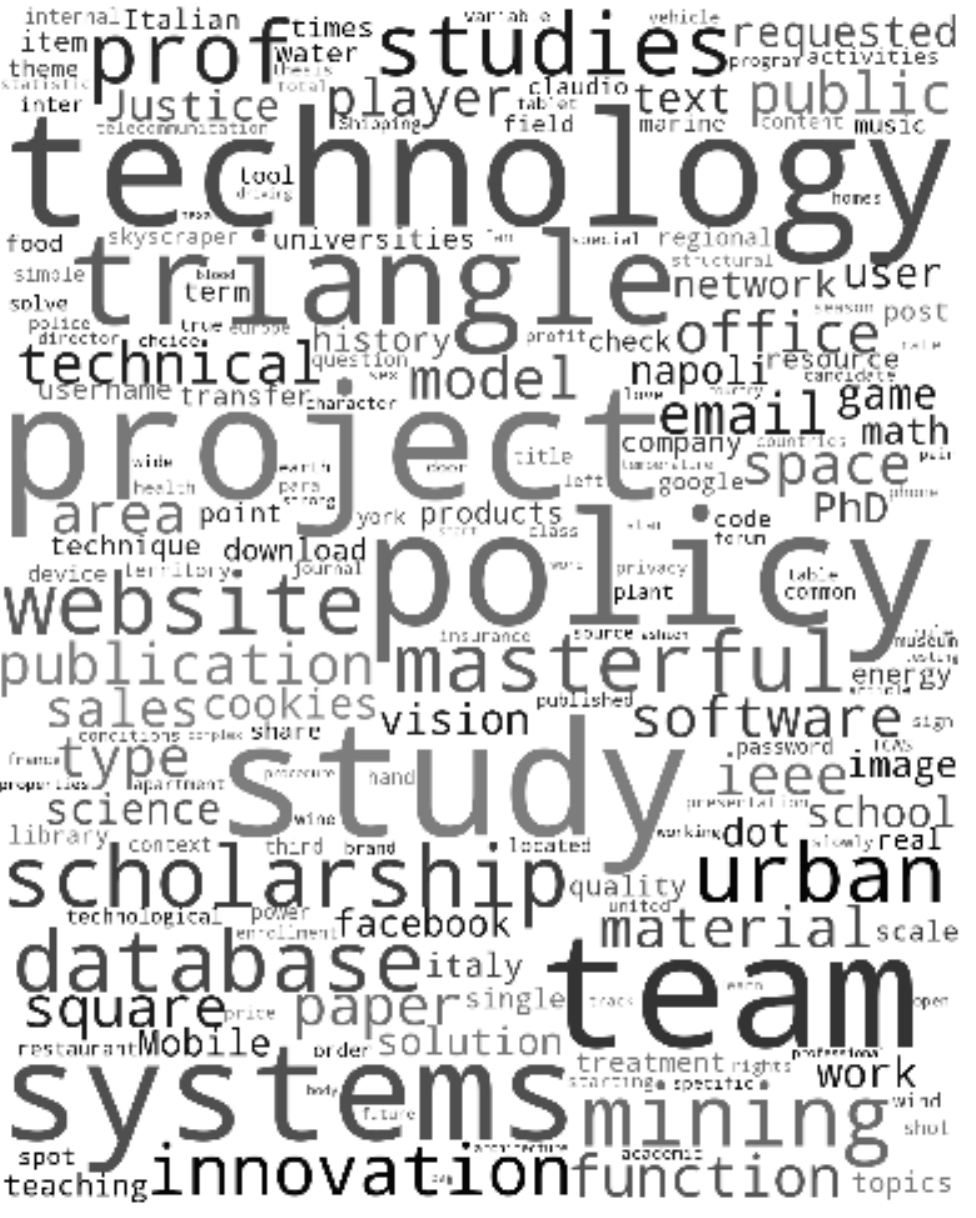}
	\caption{Wordcloud of popular keywords in the campus trace}
	\label{fig:pol_vs_pol}

\end{figure}

\begin{figure}[t!]
	 \centering
	\includegraphics[scale=0.7]{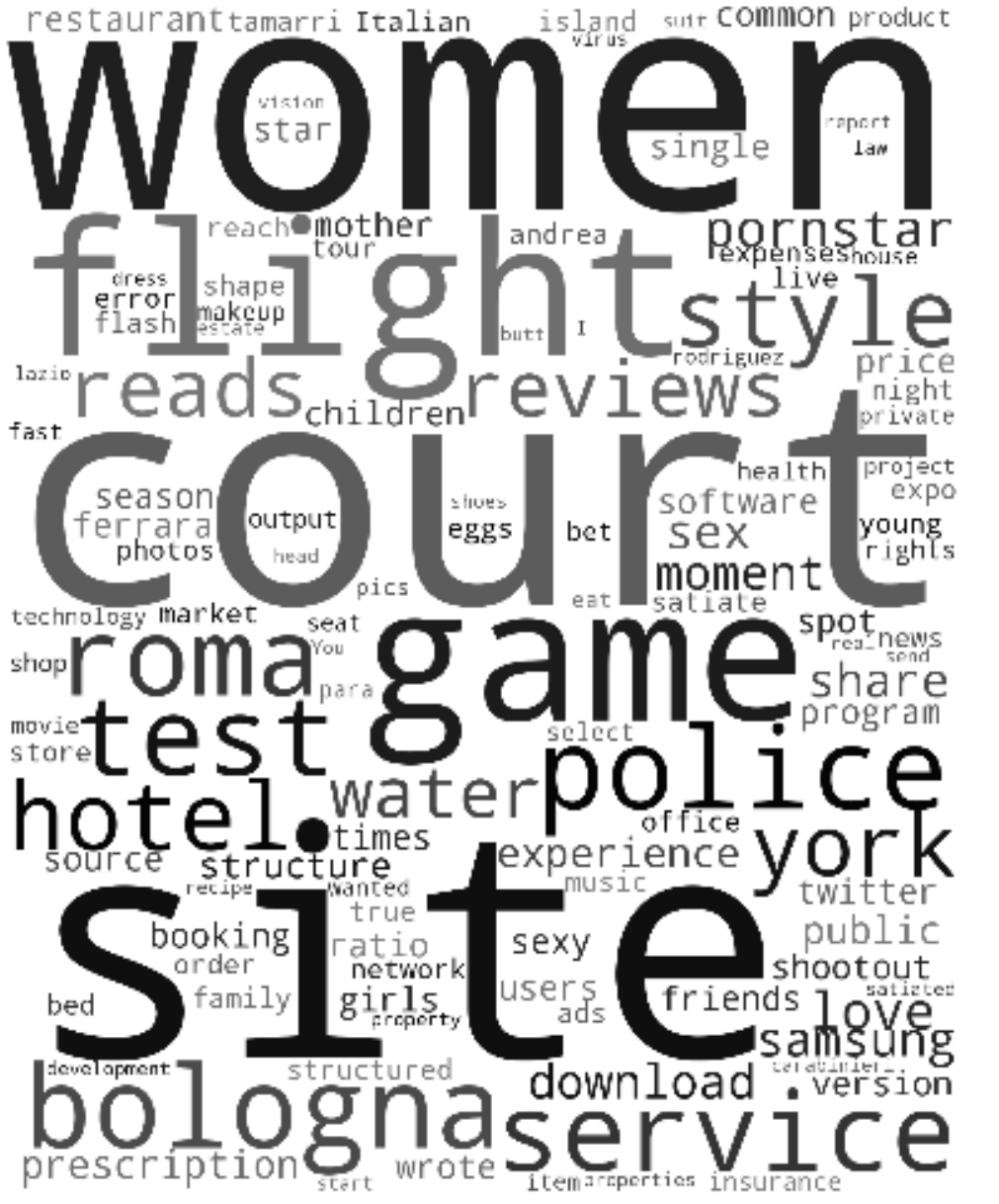}
	\caption{Wordcloud of popular keywords in a ISP trace}
	\label{fig:fast_vs_fast}

\end{figure}}
{}

\subsection{Community of a place effect}\label{sec:communities}
One of the main arguments for passive crowdsourced content promotion is its usefulness in the case of a community of a place. In this section, we characterize the community-of-a-place effect by comparing the content promoted by \HotNet in different communities. We aim to understand to what extent hot content varies across different communities of a place, and whether there is an actual community effect beyond the simple influence of the country and geographical location. To achieve this goal, we study popular topics in different communities of a place and compare them with each other. We also compare topics consumed by distinct randomly chosen populations of the same community of a place. 

We use the 1-day log traces (same day) described in Tab.~\ref{tab:desc-traces}. As shown, each trace corresponds to a different network scenario and thus community: the campus and three different ISP PoPs in the same country. 
For each trace, we extract the popular topics. First, we run \HotNet\ on the set of HTTP logs in the trace to pinpoint the corresponding content-URLs. Then, we access each content-URL, scrape the webpage and extract title and main text. For each webpage, we tokenize the text, remove stop words, and use term frequency to identify the top-10 most relevant terms, which we call \textit{keywords}.
We then weight each keyword by its overall \textit{popularity}, i.e., the sum of visits on all articles in which it appears: if a keyword appears in two pages that were viewed 3 and 5 times, the keyword's popularity is 8.
We then consider the interests of users in the trace as the set of weighted keywords.
For a fair statistical analysis, we compare keyword sets extracted from traces whose user populations have similar size. Thus, from each trace we randomly extract subsets of users so that all the populations we compare have equal size. We set the default population subset size to 3,000 users, which corresponds to the number of users we observe in our smallest vantage point. We perform different runs using different populations and average the results.


Next, we compare the keyword sets extracted from different population pairs. For a qualitative comparison, we draw a scatter plot for each pair. Dots in plots correspond to keywords, and coordinates match their popularities measured in the two populations. To quantify the correlation between keyword popularities, we measure the Pearson correlation coefficient. Note that a perfect linear relation would imply the keywords are equally popular in the two user sets, corresponding to a Pearson correlation coefficient equal to 1. Finally, as we run each comparison five times, we show one of the five scatter plots we obtain, and report the average Pearson coefficient. 
 


All in all, we test four scenarios.
In the first case, we compare the keywords we extract from distinct populations obtained by randomly picking users in \POLITOa. Fig.~\ref{fig:polito_polito} shows the results for this case. 
Notice how the dots (i.e., keywords) are distributed on the main diagonal of the graph and the Pearson coefficient is close to 1 (0.89).
This result suggests the presence of a community effect within the campus network. 
Second, we consider a residential network scenario and we compare in Fig.~\ref{fig:milano_milano} distinct populations we obtain from \HTTPtracea. Even if smaller than in the campus case, the correlation of keywords' popularity is large, testifying the presence of a community of a place effect also in this case. This is not surprising as people in the same city or neighborhood typically share common interests (e.g., local news). 
%

We consider two other scenarios. First, we compare keywords from the campus trace \POLITOa with those extracted from the ISP trace \UMB, collected at a PoP in the same city of the campus. The results, reported in Fig.~\ref{fig:polito_torino}, show a low correlation. This demonstrates that despite accessing the web from the same city, the interests observed in the two populations are different. 
\ifthenelse{\boolean{report}}
{Fig.~\ref{fig:pol_vs_pol} and Fig.~\ref{fig:fast_vs_fast} show the word clouds of popular keywords from the campus trace and from the ISP trace. We observe that those of campus trace reflect the typical activity of a university (e.g. ``study'',``scholarship'',``project''). The most popular keywords from the residential trace relate more to local services (e.g. ``court'', ``hotel'',``reviews'')}
{In fact, by building word clouds of popular keywords specific to each set, we observe that those of \POLITOa reflect the typical activity of a university, while \UMB contains many keywords related to, e.g., local news or services (see~\cite{techrep} for details).}

Finally, we compare keywords of populations belonging to two different cities. Fig.~\ref{fig:milano_torino} shows the scatter plot reporting keywords from \HTTPtracea and \HTTPtraceb. As expected, the correlation of keyword popularity is lower compared to that of two populations belonging to the same city or the same campus.

We conclude that different populations in the community of a place share common interests, confirming our intuition about the usefulness of passive content promotion in this scenario. Interestingly, our results underscore a community effect that is stronger in the campus community compared to the geographical-based one. Indeed, although people in the same neighbourhood have more common interests than people in different cities, shared interests are higher for people in the same campus. 
%
\ifthenelse{\boolean{report}}{\subsection{Complementing existing systems}\label{sec:complement}

Our user study shows that users think that \HotNet\ is simply different from the 
services they use to get informed. In this section, we study the content \HotNet\ promotes to shed light into this reasoning. We compare the frequency of keywords promoted by \HotNet\ across webpages with the frequency of those promoted by (1) a news curator (Google News), (2) an active crowdsourced system (Reddit), and, finally, (3) traditional news media. 

\noindent \textbf{\HotNet\ and Google News} Similarly to what we did in Sec.~\ref{sec:communities}, we use the keywords analysis to compare interesting topics in \HotNet and Google News. 
In particular, we focus on news media and analyze the difference between the news that Google News promotes on its homepage, and those consumed by the community of the campus and captured by \HotNet. 
On one side, we obtain a set of URLs corresponding to the news promoted by the main page of Google News for the same day when \POLITOa was collected. On the other side, we extract from \POLITOa a similar number of popular \textit{news-URLs}, i.e., content-URLs whose hostname appears in the Google News list of around 500 publishers in Italy. We remind that these news-URLs appear in both the live new stream and the Hot News sections of \HotNet. Then, for each of the two URL sets, we extract a set of keywords, and we weight each keyword with its frequency, i.e. the number of times the keyword appears across webpages (URLs).
In total, we count around 2,819 distinct keywords for \POLITOa and 2,232 for Google News. We depict in Fig.~\ref{fig:polito_gnews} the keywords with their frequencies.
Interestingly, the correlation is extremely weak, suggesting that the users in the campus are interested in different news than those promoted by Google News. This observation can be explained by two facts: first, we observe that the campus users often consume news-URLs that are one to few days old while Google News mostly targets fresh news. Second, the news promoted by Google News have country-wide interest and do not reflect the tastes of the campus users, which are however influenced by the location.

\begin{figure}[t]
  \centering
  \includegraphics[width=1.0\columnwidth]{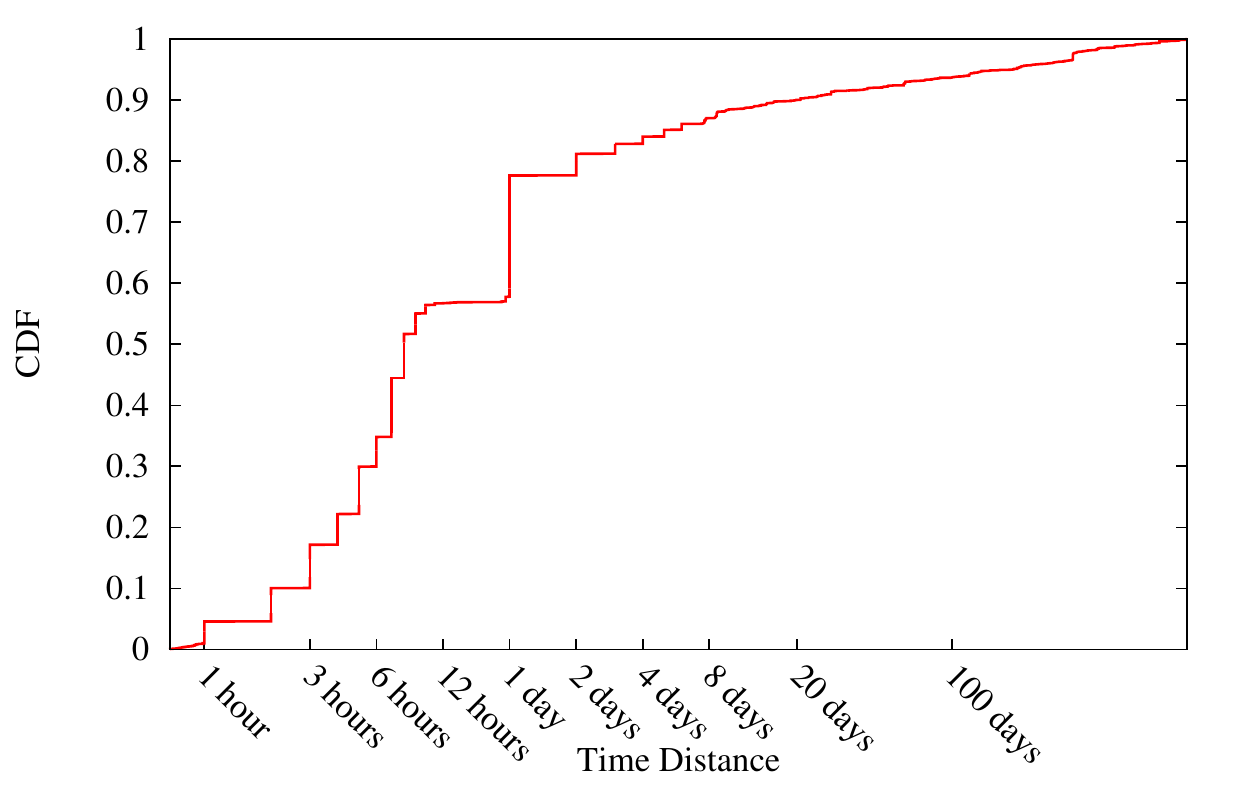}
  \caption{Freshness of news consumed by users according to Google (Online deployment).}
  \label{fig:freshness}
 \end{figure}

Fig.~\ref{fig:freshness} shows the cumulative distribution function of how long, after the Google indexing, the first {\HotNet}  user visited URLs. Note that the steps in the distribution are due to the granularity of the time information provided by Google. The figure shows that 96\% of the time, we are more than one hour behind Google news. 

Interestingly, although Google News tends to promote the most recent news (from few minutes to 2 hours) on its frontpage, what the users really view is not that fresh. Indeed, more than 44\% of news articles keep getting views from 1 to several days after their publication. A system like {\HotNet} would offer here a different perspective that is not linked to the freshness of the news, but rather to what the crowd is viewing, which is not always fresh news.\\
\noindent \textbf{\HotNet\ and Reddit} Similarly, we compare keywords extracted from the campus network (\POLITOa) with those we extract by web scraping the Italian version of Reddit's portal on the same day of the trace, considering a similar number of URLs. In this case, we observe that the correlation between the keyword sets is very weak (Pearson coefficient equal to 0.012). By manually inspecting the Reddit keywords, we observe that the topics differ from those in the campus, a large fraction of them refers to leisure activities or to Reddit itself. Moreover, we observe that the number of keywords in the campus set is much higher, which is due to a larger diversity of content. \\
\noindent \textbf{\HotNet\ and traditional news media} We perform further analysis to understand how much users rely on traditional news media for content curation.  To this end, we quantify the fraction of news-URLs we observed since the beginning of our deployment (in March 2015) and the remaining portion of content-URLs, i.e., not published by a news website in the Google News list and that we name \textit{not-news-URLs}.   We find that news-URLs represent 14\% of overall consumed content-URLs. In addition, not-news-URLs are published by a much larger set of domains (nine times larger then the list of news domains). Interestingly, in our deployment not-news-URLs tend to be often more popular than news-URLs.

}{

\noindent \textbf{\HotNet\ and Google News} Finally, since news media is one of the main sources of information in the campus community, we check if the community effect applies also to news media consumption. To this end, we use the keywords analysis to compare interesting topics in \HotNet's news and Google News. 
On one side, we obtain a set of URLs corresponding to the news promoted by the main page of Google News for the same day when \POLITOa was collected. On the other side, we extract from \POLITOa a similar number of popular \textit{news-URLs}, i.e., content-URLs whose hostname appears in the Google News list of around 500 publishers in Italy. We remind that these news-URLs appear in both the live new stream and the Hot News sections of \HotNet. Then, for each of the two URL sets, we extract a set of keywords, and we weight each keyword with its frequency, i.e. the number of times the keyword appears across webpages (URLs). In total, we count around 2,819 distinct keywords for \POLITOa and 2,232 for Google News. We depict in Fig.~\ref{fig:polito_gnews} the keywords with their frequencies. Interestingly, the correlation is extremely weak, suggesting that the users in the campus seek for different news than those promoted by Google News. This observation can be explained by two facts: first, we observe that the campus users often consume news-URLs that are one to few days old (see~\cite{techrep} for more details) while Google News mostly targets fresh news. Second, the news promoted by Google News have a country-wide interest and do not reflect the tastes of the campus users, which are influenced by the location. 
}

\vspace{1.5mm}\noindent \textbf{Content locality.}
We apply a language detection software, Google Language Detection~\cite{gld},
on the titles and text previews of the promoted content-URLs, and we find that around 94\% of them are in Italian, around 2\% are in English, and the rest covers 33 distinct languages, reflecting the diversity of the students in the campus. 
More interestingly, around 5\% of content-URLs contain in the title or text description the name of the city or the region of the campus network. Few articles even relate to the neighborhood of the university and some to the university itself. This result highlights that \HotNet\ natively offers a regional service. 


\vspace{1.5mm}\noindent \textbf{Content diversity.}
We quantify the fraction of content-URLs we observed since the beginning of our deployment (in March 2015) that were published by online newspapers. We use the Google News list of around 500 publishers in Italy (see Sec.\ref{sec:privacy_perceiving}) to split content-URLs in two parts: the first contains \textit{news-URLs}, content-URLs originated by one of these publishers; the second set contains the rest of content-URLs, that we name \textit{not-news-URLs}. We find that news-URLs represent only 14\% of overall content-URLs. In addition, not-news-URLs are published by a much larger set of domains (nine times larger then the list of news domains).  This result confirms our intuition about the fact that \HotNet\ captures a large diversity of content. Finally, in our deployment not-news-URLs are always more popular than news-URLs.

\subsection{Speed of content discovery}\label{sec:freshness}
This section measures the speed at which \HotNet discovers new Internet content, and in particular news.
A service like Google News has robots that actively look for new content to index using a predefined list of news publishers~\cite{google-news}. \HotNet, instead, relies on users exploring the web looking for fresh news. We set the bar high and leverage our campus deployment to compare the two approaches to discover news. 

To compare \HotNet to the Google News approach, each time {\HotNet} promotes a content-URL to ``Fresh news'', we check if Google has already indexed it.\footnote{We use several instances of a headless browser to do a ``site:'' search on Google Search for each news-URL {\HotNet} detects.} If so, we measure since when (this ``age'' is an information available below each link returned by Google Search). 
We ran this experiment for a period of one day (after that, Google has banned our IP network because of the extensive probing).

\begin{figure}[t!]
	 \centering
	\includegraphics[width=\columnwidth]{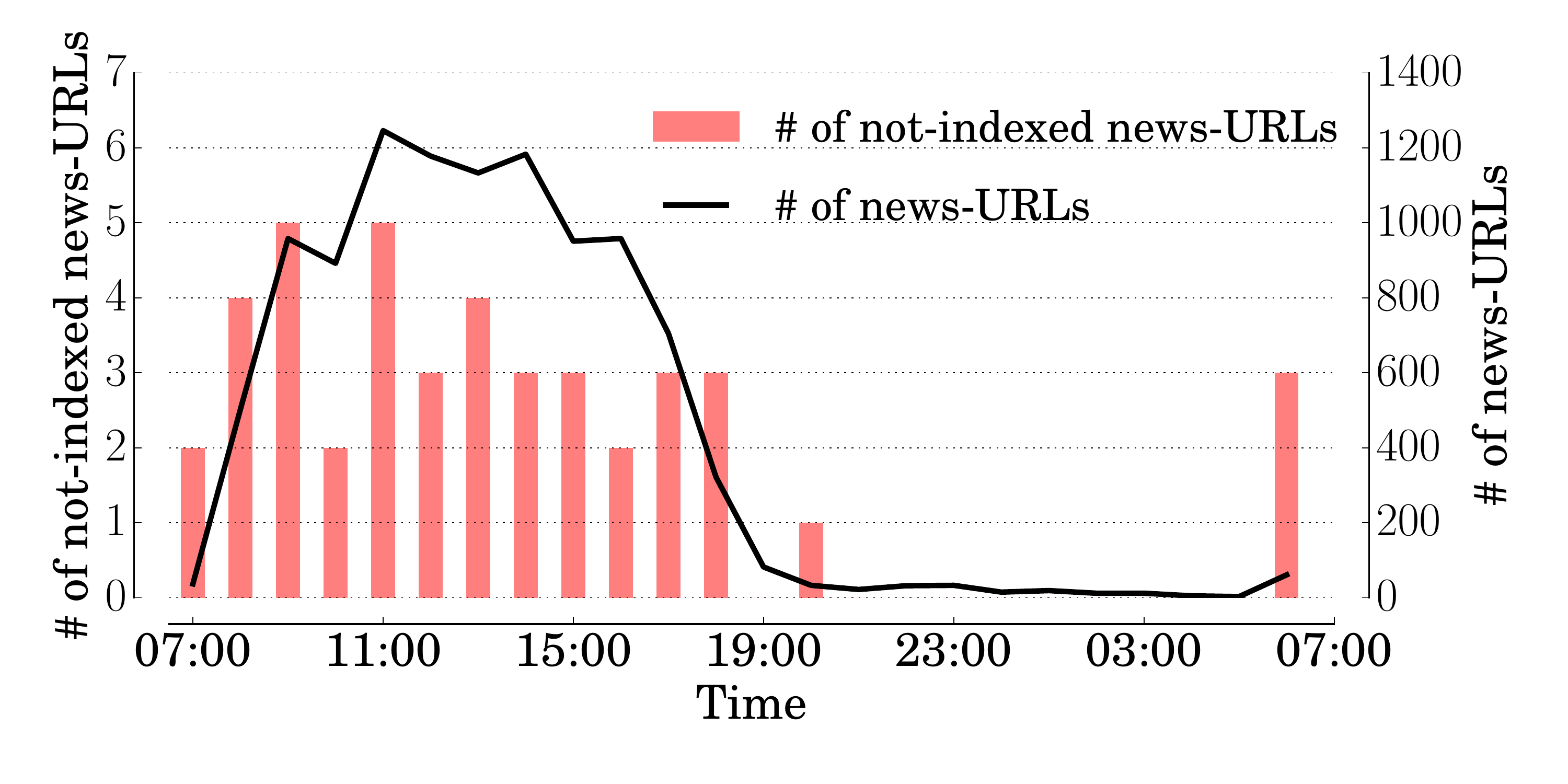}
	\vspace*{-0.8cm}
 	\caption{Number of not-indexed news-URLs (left y axis), and number of news-URLs (right y axis) during one day (Online deployment).}
 	\label{fig:notindexed}
 	\vspace{-0.4cm}
\end{figure}

Fig.~\ref{fig:notindexed} shows the number of news not indexed by Google News in each hour of the day (left-hand y axis). We put this number in perspective with the total number of ``Fresh news'' consumed per hour (right-hand y axis). The figure shows that even though we deploy \HotNet\ in a network with only a few thousand active users, it was able to find few not-yet-indexed news-URLs. Not surprisingly, there is a positive correlation between the number of viewed news-URLs per hour and the number of news-URLs in \HotNet\ that have not yet been indexed by Google News.

We do not expect \HotNet\ to compete with a large-scale active crawling system like Google News on the speed of news discovery. But clearly, increasing the number of users would increase the speed of content discovery. This would mean deploying passive crowdsourced content curation in a setting that goes beyond the community of a place scenario we explore in this paper. However, popular toolbars, a search engine like Google or a social network like Facebook have however access to a much larger number of clicks and could try implementing this.


 
\vspace*{-0.3cm}
\section{Related Work}\label{sec:related}

The research community has spent a considerable effort on studying community-based promotion mechanisms. Notable examples like~\cite{quijano2013social,sharma2013social,amin2012social,guy2009personalized} demonstrate that in general social-based recommendation leads to better results than recommendation based on users' preference similarity, in an heterogeneous set of taste-related domains (e.g., movie catalogs, club ratings, webpages and blogs).
Based on these observations, some systems have been designed specifically for content promotion in communities-of-a-place scenarios such as campuses and neighborhoods. For example, several enterprises and universities deploy internal social networks~\cite{dimicco2008motivations,hideyuki2007innovationcafe,du2012university}, create blogs~\cite{kim2008investigating} and wikis~\cite{hasan2006wiki}.

Even if not suitable for content recommendation in communities of a place, Reddit, Storify and Pearltrees are crowdsourced content sharing platforms which build on a community to vote and promote contents. Although powerful, prior work highlighted some of their weaknesses. First, most users participating in platforms like Reddit are not active. This slows down information propagation in the community~\cite{hotnet:pullrec}. Second, although they target the web at large, such communities tend to become closed and self-referencing, with mostly user-generated content~\cite{reddit:selfref}. Third, they heavily suffer from ``freeloaders'', i.e., passive users relying on others actively sharing contents, and from a large amount of submitted contents which are not viewed~\cite{reddit:underprovision}. Other tools like Facebook's Paper~\cite{facebook-paper} and Twitter Trends~\cite{twitter-trends} overcome these issues as they build on huge social networks to discover novel and popular content. Similarly, social bookmarking tools like Delicious and Pinterest, promote content based on users who organize collections of webpages.  

All above platforms share the problem of requiring users to be active at sharing content, which is difficult to obtain, in general~\cite{friedman2014enterprise,mcafee2006emergent}.
Differently, \HotNet\ is a system for content sharing in communities of a place based solely on the passive observation of network traffic, and, to the best of our knowledge, it is the first system in its kind.
\HotNet's most similar system is perhaps Google Trends~\cite{google-trends}, which passively observes users' activity in Google Search (i.e., search queries) to extract trends about the most searched topics. The result however is different from the one produced by \HotNet, as Google Trends returns the keys used for search queries, instead of links to contents.

Other proposals aim at personalizing the recommended contents to match users' interests~\cite{Liu:2010:google:pers}, or to offer news based on a regional basis~\cite{Liu:2008:LAL}. \HotNet, in its current design does not offer personalized recommendations beyond the community-based aspect, but we plan to explore this aspect in future work.

Finally, \HotNet is not the first system tackling the task of mining HTTP logs for URL extraction. For instance, this idea was already proposed by authors of~\cite{Jia_ICDM07}. However, the result obtained with their approach, i.e., number of clicks to web portals, is not different from what services like Alexa already provide.
Indeed, the main challenge behind this task resides in the detection of what we call user-URLs from the set of all URLs, and only a few other research studies address this problem~\cite{time-based,type-based,IMC-2011-referrer,xie-resurf-2013}. Among them, only StreamStructure~\cite{IMC-2011-referrer} and ReSurf~\cite{xie-resurf-2013} are still adequate to investigate the complex structure of modern web pages. Neither of them, however, are appropriate for \HotNet, as they rely on HTTP Content-Type, which is often unreliable~\cite{DBLP:conf/pam/SchneiderAMFU12}, and makes traffic extraction more complex (because of the matching between requests and responses) and unsuitable for an online system as \HotNet. Our experiments in Sec.~\ref{sec:online:perf} confirm this observation. Our preliminary work~\cite{Houidi_TMA14} designed the offline version of the algorithm we use to detect user-URLs (Sec.~\ref{sec-filtering-browser}). In this paper we present an online version of this algorithm, we evaluate its accuracy, and compare with ReSurf on the same ground-truth dataset.
\section{Ethical and legal issues}\label{sec:ethical_and_legal}

Deploying a system like \HotNet\ comes with a series of ethical and legal issues. 

First of all, the users accessing the Web from the monitored network might suffer from the ``Big Brother effect'', i.e., experiencing the fear of having someone spying on their browsing activity. Surprisingly, however, after the first announcement of \HotNet, only two people expressed some privacy worries, an union official and one respondent to the evaluation form. 
In the subsequent announcements and talks presenting \HotNet, we took care of describing the privacy-preserving mechanisms \HotNet adopts, and highlighted that these were approved by the Security and Privacy boards of Politecnico di Torino, Inria, Nokia and the ISP collaborating with us. Moreover, we enriched the website with a section describing these mechanisms.

Finally, \HotNet might promote URLs pointing to inappropriate or potentially damaging content (e.g., porn or malwares). Also, the promoted content may expose internal information that the deploying institution is not willing to share. Although, we have not faced any of these issues yet, we plan to allow the deploying institution to create blacklists to filter out unwelcome hostnames and users to notify contents they consider inappropriate or offending.

\vspace*{-0.2cm}
\section{Conclusion}\label{conclusion}
This paper designed, implemented and evaluated \HotNet, the first passive crowdsourced system based on web-clicks to boost content discovery in communities of a place. 
\HotNet takes as input a stream of HTTP requests observed in a community and processes it online to detect URLs visited by users, and uses different methods to promote the most interesting ones while preserving user privacy. Our system evaluation with real and ground-truth HTTP traces showed that our algorithms are accurate and efficient in processing, online, HTTP logs at scale. 
We have deployed \HotNet in a large campus network and in an ISP neighbourhood, and used these deployments  for evaluation. Two other deployments in a research institute and a corporate network are underway. 
Of the campus users who evaluated \HotNet, more than 90\% welcome the quality of \HotNet\ content. 
Our analysis of \HotNet's liveness shows that \HotNet becomes interesting starting from around 1000 active users browsing the web. Finally, our analysis of the topics observed in different communities of a place confirm the promise of passive crowdsourcing to foster content discovery in such under-engaged communities, and show that it nicely complements existing systems. 

We believe the approach we propose for content promotion could be inspiring for big players of the web who have access to clicks from large user populations. A large scale variation of \HotNet could be an interesting approach to reduce filter bubble effects and expose users to a larger variety of content. 
\bibliography{report-bib}

\begin{thebibliography}{10}

\bibitem{ManagementInformationSystems}
K.~C. Laudon and J.~P. Laudon, {\em Management information systems}, vol.~6.
\newblock Prentice Hall Upper Saddle River, NJ, 2000.

\bibitem{sellen2002knowledge}
A.~J. Sellen, R.~Murphy, and K.~L. Shaw, ``How knowledge workers use the web,''
  in {\em ACM SIGCHI}, 2002.

\bibitem{Kwak:2010:twitter}
H.~Kwak, C.~Lee, H.~Park, and S.~Moon, ``{What is Twitter, a social network or
  a news media?},'' in {\em ACM WWW}, 2010.

\bibitem{mcafee2006enterprise}
A.~P. McAfee, ``Enterprise 2.0: The dawn of emergent collaboration,'' {\em MIT
  Sloan management review}, vol.~47, no.~3, 2006.

\bibitem{friedman2014enterprise}
B.~D. Friedman, M.~J. Burns, and J.~Cao, ``Enterprise social networking data
  analytics within alcatel-lucent,'' {\em Bell Labs Technical Journal},
  vol.~18, no.~4, 2014.

\bibitem{GilbertCSCW13}
E.~Gilbert, ``{Widespread Underprovision on Reddit},'' in {\em ACM CSCW}, 2013.

\bibitem{digg-patriots}
D.~Neal, ``Digg is dogged by conservative pressure groups.''
  \url{http://goo.gl/M9Plt}.
\newblock [Online; August 2010].

\bibitem{passcrowd}
``{Passive Crowdsourcing: 5 Ways We Do Work for Others Without Realizing It}.''
  \url{http://goo.gl/0GEW9O}.

\bibitem{reddit-stats}
``Reddit statistics.'' \url{https://www.reddit.com/about/}.
\newblock Accessed February 2, 2016.

\bibitem{Naylor2014}
D.~Naylor, A.~Finamore, I.~Leontiadis, Y.~Grunenberger, M.~Mellia,
  M.~Munaf\`{o}, K.~Papagiannaki, and P.~Steenkiste, ``{The Cost of the "S" in
  HTTPS},'' in {\em ACM CoNext}, 2014.

\bibitem{finamore2011_tstat}
A.~Finamore, M.~Mellia, M.~Meo, M.~M. Munaf{\`o}, and D.~Rossi, ``{Experiences
  of {I}nternet traffic monitoring with {T}stat},'' {\em IEEE Network}, 2011.

\bibitem{fraleigh:03}
C.~Fraleigh, S.~Moon, B.~Lyles, C.~Cotton, M.~Khan, R.~Rockell, D.~Moll,
  T.~Seely, and C.~Diot, ``{Packet-level traffic measurements from the Sprint
  IP backbone},'' in {\em IEEE Network Magazine}, vol.~17, Nov 2003.

\bibitem{IMC-2011-referrer}
S.~Ihm and V.~S. Pai, ``{Towards Understanding Modern Web Traffic},'' in {\em
  ACM IMC}, 2011.

\bibitem{xie-resurf-2013}
G.~Xie, M.~Iliofotou, T.~Karagiannis, M.~Faloutsos, and Y.~Jin, ``{ReSurf:
  Reconstructing web-surfing activity from network traffic},'' in {\em IFIP
  Networking}, 2013.

\bibitem{cryptopan}
``{Crypto-PAn}.'' \url{http://goo.gl/tru2ti}.

\bibitem{ButToN2013}
M.~Butkiewicz, H.~Madhyastha, and V.~Sekar, ``{Characterizing Web Page
  Complexity and Its Impact},'' {\em IEEE/ACM ToN}, 2013.

\bibitem{Houidi_TMA14}
Z.~Ben~Houidi, G.~Scavo, S.~Ghamri-Doudane, A.~Finamore, S.~Traverso, and
  M.~Mellia, ``Gold mining in a river of internet content traffic,'' in {\em
  TMA}, 2014.

\bibitem{ad-block}
``{Adblock Plus}.'' \url{http://easylist.adblockplus.org/}.
\newblock [Online; Feb 2016].

\bibitem{class-imbalance}
Y.~Sun, A.~K.~C. Wong, and M.~S. Kamel, ``Classification of imbalanced data: a
  review,'' {\em International Journal of Pattern Recognition and Artificial
  Intelligence}, 2009.

\bibitem{reddit-hot}
``{R}eddit's source code".'' \url{https://github.com/iangreenleaf/reddit}.

\bibitem{techrep}
G.~Scavo, Z.~Ben~Houidi, S.~Traverso, R.~Texeira, and M.~Mellia, ``{Technical
  report: Mining HTTP logs online for network-based content recommendation },''
  \url{http://www.retitlc.polito.it/traverso/papers/webrowse_tech-rep.pdf}.

\bibitem{surveyrate}
``{Survey Response Rates}.'' \url{http://goo.gl/NiW471}.

\bibitem{gld}
``{Google Language Detection}.''
  \url{http://code.google.com/p/language-detection/}.

\bibitem{google-news}
Google, ``Google news.'' \url{https://news.google.com/}.
\newblock Accessed May 6, 2015.

\bibitem{quijano2013social}
L.~Quijano-Sanchez, J.~A. Recio-Garcia, B.~Diaz-Agudo, and G.~Jimenez-Diaz,
  ``Social factors in group recommender systems,'' {\em ACM TIST}, 2013.

\bibitem{sharma2013social}
A.~Sharma and D.~Cosley, ``Do social explanations work?: studying and modeling
  the effects of social explanations in recommender systems,'' in {\em ACM
  WWW}, 2013.

\bibitem{amin2012social}
M.~S. Amin, B.~Yan, S.~Sriram, A.~Bhasin, and C.~Posse, ``Social referral:
  leveraging network connections to deliver recommendations,'' in {\em ACM
  RecSys}, 2012.

\bibitem{guy2009personalized}
I.~Guy, N.~Zwerdling, D.~Carmel, I.~Ronen, E.~Uziel, S.~Yogev, and
  S.~Ofek-Koifman, ``Personalized recommendation of social software items based
  on social relations,'' in {\em ACM RecSys}, 2009.

\bibitem{dimicco2008motivations}
J.~DiMicco, D.~R. Millen, W.~Geyer, C.~Dugan, B.~Brownholtz, and M.~Muller,
  ``Motivations for social networking at work,'' in {\em ACM CSCW}, 2008.

\bibitem{hideyuki2007innovationcafe}
F.~Hideyuki, ``{Innovationcafe an In-house Social Network Service (SNS) Used in
  NEC},'' {\em NEC Technical Journal}, vol.~2, no.~2, 2007.

\bibitem{du2012university}
Z.~Du, X.~Fu, C.~Zhao, and T.~Liu, ``University campus social network system
  for knowledge sharing,'' in {\em IEEE ICSAI}, 2012.

\bibitem{kim2008investigating}
S.~T. Kim, C.~K. Lee, and T.~Hwang, ``Investigating the influence of employee
  blogging on it workers' organisational citizenship behaviour,'' {\em
  International Journal of Information Technology and Management}, 2008.

\bibitem{hasan2006wiki}
H.~Hasan and C.~C. Pfaff, ``The wiki: an environment to revolutionise
  employees' interaction with corporate knowledge,'' in {\em ACM OZCHI}, 2006.

\bibitem{hotnet:pullrec}
H.~V. Madhyastha and M.~Maiya, ``Towards comprehensive social sharing of
  recommendations: augmenting push with pull,'' in {\em ACM HotNets}, 2013.

\bibitem{reddit:selfref}
P.~Singer, F.~Fl{\"o}ck, C.~Meinhart, E.~Zeitfogel, and M.~Strohmaier,
  ``{Evolution of Reddit: from the front page of the Internet to a
  self-referential community?},'' in {\em ACM WWW}, 2014.

\bibitem{reddit:underprovision}
E.~Gilbert, ``Widespread underprovision on reddit,'' in {\em Conference on
  Computer supported cooperative work}, ACM, 2013.

\bibitem{facebook-paper}
``Facebook paper.'' \url{https://www.facebook.com/paper}.

\bibitem{twitter-trends}
``Twitter trends/place api.'' \url{https://goo.gl/GQ776w}.

\bibitem{mcafee2006emergent}
A.~P. McAfee, ``Emergent collaboration,'' 2006.

\bibitem{google-trends}
``Google trends.'' \url{http://www.google.com/trends/}.

\bibitem{Liu:2010:google:pers}
J.~Liu, P.~Dolan, and E.~R. Pedersen, ``Personalized news recommendation based
  on click behavior,'' in {\em ACM IUI}, 2010.

\bibitem{Liu:2008:LAL}
J.~Liu and L.~Birnbaum, ``Localsavvy: aggregating local points of view about
  news issues,'' in {\em ACM LOCWEB}, 2008.

\bibitem{Jia_ICDM07}
M.~Jia, S.~Ye, X.~Li, and J.~Dickerson, ``Web site recommendation using http
  traffic,'' in {\em IEEE ICDM}, 2007.

\bibitem{time-based}
P.~Barford and M.~Crovella., ``Generating representative web workloads for
  network and server performance evaluation,'' in {\em ACM SIGMETRICS}, 1998.

\bibitem{type-based}
H.-K. Choi and J.~O. Limb, ``A behavioral model of web traffic,'' in {\em IEEE
  ICNP}, 1999.

\bibitem{DBLP:conf/pam/SchneiderAMFU12}
F.~Schneider, B.~Ager, G.~Maier, A.~Feldmann, and S.~Uhlig, ``Pitfalls in http
  traffic measurements and analysis,'' in {\em PAM}, 2012.

\end{thebibliography}
\bibliographystyle{ieeetr}
\end{document}